# Depletion interactions modulate coupled folding and binding in crowded environments

Franziska Zosel[1,2], Andrea Soranno[1,3], Daniel Nettels[1], & Benjamin Schuler[1,4]

[1]Department of Biochemistry, University of Zurich, Winterthurerstrasse 190, 8057 Zurich, Switzerland
[2]Current Address: Novo Nordisk A/S, Novo Nordisk Park, 2760 Måløv, Denmark
[3]Department of Biochemistry and Molecular Biophysics, Washington University in St. Louis, Missouri 63130, United States
[4]Department of Physics, University of Zurich, Winterthurerstrasse 190, 8057 Zurich, Switzerland

Correspondence should be addressed to F.Z. (franziska.zosel@gmail.com), A.S. (soranno@wustl.edu), or B.S. (schuler@bioc.uzh.ch).

## Abstract

Intrinsically disordered proteins (IDPs) abound in cellular regulation. Their interactions are often transitory and highly sensitive to salt concentration and posttranslational modifications. However, little is known about the effect of macromolecular crowding on the kinetics and stability of the interactions of IDPs with their cellular targets. Here, we investigate the influence of crowding on the coupled folding and binding between two IDPs, using polyethylene glycol as a crowding agent across a broad size range. Single-molecule Förster resonance energy transfer allows us to quantify several key parameters simultaneously: equilibrium dissociation constants, kinetic association and dissociation rates, and translational diffusion coefficients resulting from changes in microviscosity. We find that the stability of the IDP complex increases not only with the concentration but also with the size of the crowders, in contradiction to scaled-particle theory. However, both the equilibrium and the kinetic results can be explained quantitatively by depletion interactions if the polymeric properties of proteins and crowders are taken into account. This approach thus provides an integrated framework for addressing the complex interplay between depletion effects and polymer physics on IDP interactions in a crowded environment.

## Significance Statement

The molecular environment in a biological cell is much more crowded than the conditions commonly used in biochemical and biophysical experiments *in vitro*. It is therefore important to understand how the conformations and interactions of biological macromolecules are affected by such crowding. Addressing these questions quantitatively, however, has been challenging owing to a lack of sufficiently detailed experimental information and theoretical concepts suitable for describing crowding, especially when polymeric crowding agents and biomolecules are involved. Here, we use the combination of extensive single-molecule experiments with theoretical concepts from soft-matter physics to investigate the interaction between two intrinsically disordered proteins. We observe pronounced effects of crowding on their interactions and succeed in providing a quantitative framework for rationalizing these effects.

## Introduction

Intrinsic disorder is a widespread phenomenon among eukaryotic proteins, manifesting itself in unstructured segments of larger proteins or proteins that are entirely disordered under physiological conditions[1]. Such intrinsically disordered proteins (IDPs) are particularly prevalent in the context of signaling and regulation[2], where they form complex interaction networks[3], often involving many partners[4]. IDPs lack the stable tertiary structure familiar from folded proteins – instead, they sample a heterogeneous ensemble of conformations on timescales from nanoseconds to seconds[5-8]. Their disorder and the lack of pronounced minima in their conformational free energy makes the ensembles particularly sensitive to external factors such as ligands[9], posttranslational modifications[10], and salt concentration[11], which can even induce the folding of IDPs. The cellular milieu, densely packed with globular and polymeric macromolecules[12-14], is thus also expected to influence the conformational distributions and dynamics of IDPs. Experimental evidence, simulations, and theory suggest that the effects of such macromolecular crowding are moderate but detectable, including the compaction and local structure formation of unfolded and intrinsically disordered proteins[8,15-32], which may have important effects on their function. However, a quantitative understanding of the effects of crowding on IDPs is largely lacking.

While the influence of macromolecular crowding on the conformational properties of individual IDPs and on the binding interactions of folded proteins has been studied[33-36], little is known about how crowding affects the process of coupled folding and binding, which is involved in many functional interactions of IDPs[9,37]. Here, we aim to fill this gap with a systematic investigation of the effects of the size and concentration of polymeric crowding agents on the coupled folding and binding of two IDPs by simultaneously monitoring complex stability, kinetics, and translational diffusion. We find that all of these aspects can be rationalized quantitatively within the framework of depletion interactions[38-40], which allows us to combine the influence of polymer physics[19] with the enhanced attractive interactions between the proteins in a crowded solution[40].

## Results

### Probing coupled folding and binding with single-molecule FRET

We investigate the interaction between the intrinsically disordered activation domain of the steroid receptor coactivator 3 (ACTR) and the molten-globule-like nuclear coactivator binding domain of CBP/p300 (NCBD), a paradigm of coupled folding and binding[41,42]. Upon binding to each other, ACTR and NCBD form a stable, structured complex[41] with an equilibrium dissociation constant of ~25 nM (Fig. 1$A$). Association is fast (~$10^8$ M$^{-1}$s$^{-1}$) and electrostatically favored by the opposite net charge of the two proteins[7,43,44]. We monitor the binding of NCBD to surface-immobilized ACTR molecules in confocal single-molecule FRET experiments (Fig. 1$B$)[7]. To follow the binding reaction, we labeled ACTR on its C-terminus with a fluorescent donor dye, and NCBD on its N-terminus with a fluorescent acceptor dye. In the unbound state of ACTR, only donor fluorescence is observed (with some background in the acceptor channel from freely diffusing NCBD). Upon binding, the donor and acceptor dyes of ACTR and NCBD come into proximity, resulting in Förster resonance energy transfer (FRET) between them, as evident from the increase in acceptor intensity and simultaneous decrease in donor intensity. When NCBD dissociates, acceptor emission ceases, and the donor fluorescence returns to its original intensity, leading to anticorrelated signal changes of donor and acceptor (Fig. 1$B$).

Fig. 1 illustrates that each such measurement enables us to acquire a comprehensive set of observables. From the fluorescence time traces, both the equilibrium dissociation constant, $K_D$ (Fig. 1$C$), and the kinetic on- and off-rate coefficients of the binding reaction, $k_{on}$ and $k_{off}$ (Fig. 1$D$), can be quantified (see Materials and Methods). From fluorescence correlation spectroscopy (FCS) measurements in the solution above the surface (Fig. 1*E*), we can further determine the diffusion time, $\tau_D$, of acceptor-labeled NCBD through the confocal volume of the instrument to quantify the translational diffusion coefficient, *D*. Finally, FCS also reports on the average number of molecules in the confocal volume via its amplitude, which allows us to correct for small variations in NCBD concentration[*]. The complementarity of these observables, all of which are obtained under identical solution conditions, enables an integrated analysis of the effects of crowding on the coupled folding and binding reaction.

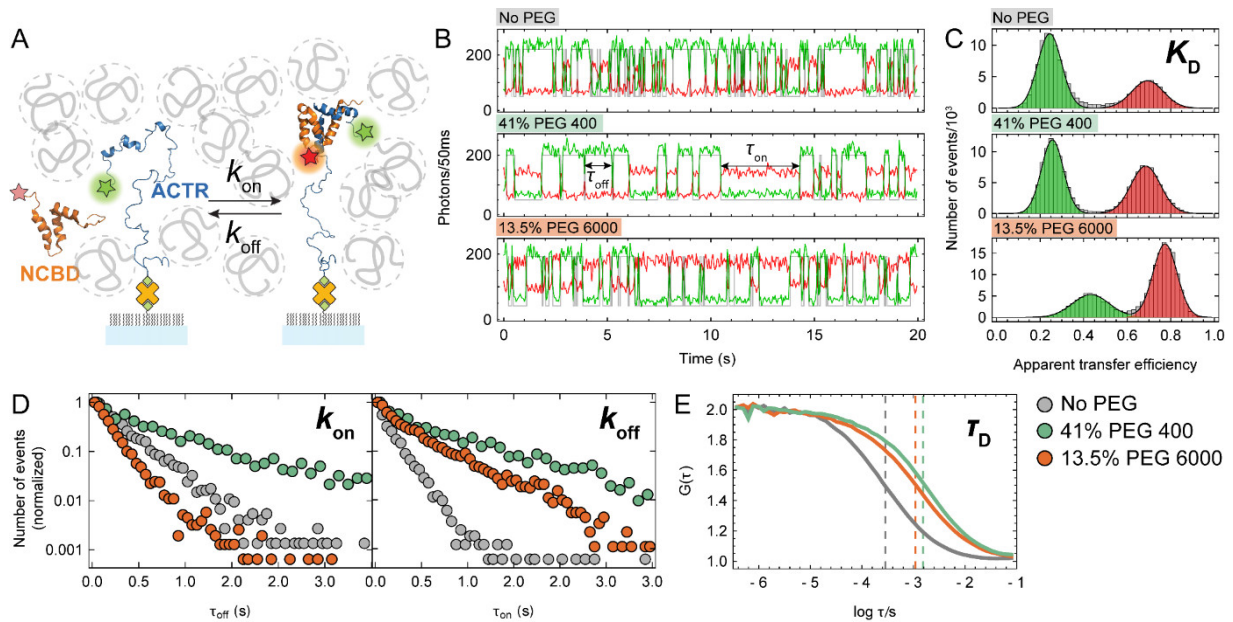

**Figure 1. Single-molecule experiments reveal comprehensive information on crowding effects.** (*A*) Schematic representation of acceptor-labeled NCBD (orange, PDB entry 2KKJ) binding to surface-immobilized donor-labeled ACTR (blue) in the presence of polymeric crowders (gray). (*B*) Examples of single-molecule time traces recorded at different PEG concentrations (first 20 seconds each, binning: 50 ms, donor signal: green, acceptor signal: red; not corrected for background, quantum yields, detection efficiencies, etc.). The most likely state trajectory identified by the Viterbi algorithm is depicted in gray. *From top to bottom*: buffer without PEG; 36% PEG 400; 13.5% PEG 6000. (*C*) Histograms of the apparent transfer efficiency (from time traces binned at 20 ms) can be used to quantify the equilibrium dissociation constant, $K_D$. Apparent efficiencies at 13.5% PEG 6000 are shifted to higher values because of the increased background in the acceptor channel owing to residual nonspecific surface adsorption of NCBD. (*D*) Normalized dwell-time distributions (conditions and color code as in *B*) from the state trajectories of 30-40 ACTR molecules each (gray: 9300 transitions, green: 3475 transitions, orange: 12192 transitions). The dwell time distributions in the unbound and bound states yield $k_{on}$ (*left panel*) and $k_{off}$ (right panel), respectively (see Methods). (*E*) Normalized FCS curves of freely diffusing acceptor-labeled NCBD measured above the surface under the same conditions as the time traces (*B*) yield the diffusion times of NCBD through the confocal volume (dashed lines), which can be related to translational diffusion coefficients.

---

[*] In spite of surface passivation, small variations in the exceedingly low NCBD concentrations from measurement to measurement can result from loss of sample by adsorption of NCBD to the surface of the cover slide or sample chamber, especially in solutions containing high concentrations of large PEGs (cf. Fig. S1).

As crowding agents (crowders), we chose (poly)ethylene glycol (PEG) because it is widely used for mimicking inert crowders[45,46]; its interaction with proteins is dominated by excluded-volume effects (especially for longer PEG chains)[47,48]; and it is available over a wide range of degrees of polymerization at a purity suitable for single-molecule fluorescence experiments, even at physiologically realistic mass fractions of up to ~40%.[19] We can thus investigate a large range of relative protein-crowder dimensions, including crowders that are much smaller, of similar size, and much larger than the proteins used. NCBD and ACTR have hydrodynamic radii of $R_H$ = 1.74 nm and $R_H$ = 2.3 nm, respectively, as determined by NMR[49] and 2-focus FCS[50], so we selected ten different degrees of polymerization, $P$, of PEG (Fig. S1$A$,$B$), ranging from the monomer, ethylene glycol ($R_g$ ≈ 0.2 nm), to PEG 35000 ($R_g$ ≈ 10 nm). For every set of solution conditions (71 in total), 34 to 83 min of cumulative single-molecule time traces were analyzed, each set corresponding to $3 \cdot 10^3$ to $2.5 \cdot 10^4$ association/dissociation transitions (Table S1), to enable a comprehensive quantitative analysis.

It is worth noting that we recently found that NCBD exists in two conformations corresponding to different peptidyl-prolyl cis/trans isomers, both of which are able to bind ACTR, but with different affinities and dissociation rates[7]. The relative effects of crowders on the kinetics and affinities of both isomers is, however, equal to within experimental error (Fig. S2, Table S1). For the sake of clarity, we thus focus on the simpler two-state analysis here. The parameters extracted for each set of conditions are compiled in Table S1.

**Polymeric crowders and relevant length scales**

For investigating the effects of a polymeric crowder, it is essential to recognize that we do not only cover a large range of PEG sizes but also two different concentration regimes (Fig. S1). At low polymer concentrations, in the dilute regime, the sizes of the polymeric crowders can be approximated by their radii of gyration, $R_g$, since the chains do not overlap.[51] With increasing crowder concentration, the chains fill the available volume more and more, until they start to overlap, at which point the solution enters the semidilute regime. The overlap concentration, $c^*$, separating the two regimes (used here in units of mass per volume[40]) is given by

$$c^* = M \Big/ \left( N_A \tfrac{4}{3} \pi R_g^3 \right) \propto P^{-0.749}, \tag{1}$$

where $M$ is the molar mass of the crowder and $N_A$ Avogadro's constant. The overlap concentration of PEG thus strongly decreases with increasing $P$ (Fig. S1$C$). Within the accessible range of $c$, we reach the semidilute regime for PEGs with $M ≥ 1000$ g/mol; for $M ≥ 4600$ g/mol, most of the recorded data points are in the semidilute regime (Fig. S1$C$).[†] In the semidilute regime, the characteristic length scale is no longer $R_g$ of the individual polymer chains, but the average mesh size, $\xi$, in the network of overlapping polymers. In this sense, the solution can also be viewed as a solution of 'blobs' of size $\xi$. Inside a blob, the monomers of a chain do not overlap with other chains, whereas on length scales greater than a blob (or correlation length), the excluded volume interactions within the protein and within the crowding agents are screened by other overlapping chains[52]. Importantly, $\xi$ is independent of $P$ but decreases steeply with increasing $c$ (see Theory)[40,52].

---

[†] The upper limit in the PEG concentrations usable for PEGs with $M ≥ 1000$ g/mol was given by nonspecific surface adsorption of NCBD, which causes a high background signal in the acceptor channel and interferes with data analysis (see also Fig. 1$B$, *lower panel*).

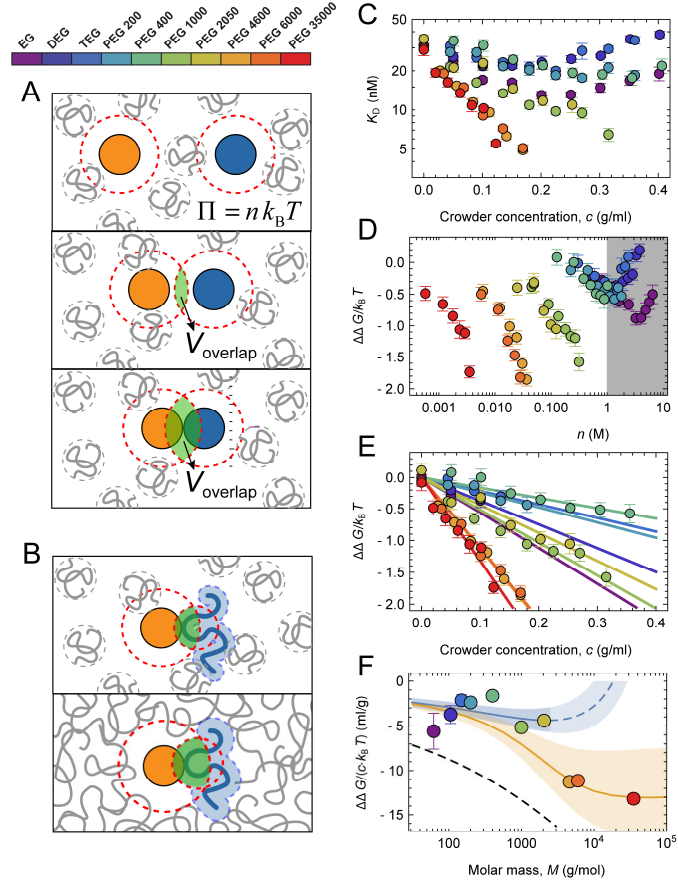

**Figure 2. Depletion interactions stabilize the IDP complex.** (*A*) Depletion interaction between two spherical colloidal particles in a solution of non-interacting polymers[38,39]. Each particle has a depletion layer into which the centers of mass of the polymers cannot enter. When the depletion layers of the two particles overlap, the volume available to the polymer chains increases by the overlap volume $V_{\text{overlap}}$ (green), which increases their entropy and causes an attractive potential between the two particles via the osmotic pressure, $\Pi = n k_{\text{B}} T$. (*B*) The theory is modified to account for the smaller overlap volume between the compact, molten-globule-like protein NCBD and the largely unstructured IDP ACTR, whose depletion layer arises from the size of the interacting segment instead of $R_{\text{H}}$ of the whole protein. (*C*) Equilibrium dissociation constant, $K_{\text{D}}$, for the interaction between ACTR and NCBD, $K_{\text{D}} = k_{\text{off}} / k_{\text{on}}$, as a function of PEG concentration, $c$. (*D*) Change in interaction free energy between ACTR and NCBD caused by crowding, $\Delta\Delta G/k_{\text{B}}T$ (Eq. (4)), versus the number concentration of PEG, $n$. Data for $n > 1$ M (shaded range) were excluded from the analysis. (*E*) Linear fit of $\Delta\Delta G/k_{\text{B}}T$ as a function of $c$ for $n \leq 1$ M. (*F*) Magnitude of the crowder concentration dependence, $\Delta\Delta G/ck_{\text{B}}T$, i.e., the slope from the linear fit in *E*, as a function of crowder size. The black dashed line indicates the dilute-limit prediction for two spherical particles with $R$ = 1.74 and 2.3 nm, the measured $R_{\text{H}}$ values of NCBD and ACTR (Eq. (4)). The blue line shows the dilute-limit prediction using instead a segment size of 0.4 ± 0.1 nm for ACTR (Eq. (4)). The orange line shows the corresponding prediction in the semidilute regime[40], where the overlap volume is determined by the correlation length $\xi$, and the osmotic pressure is approximated by renormalization group theory (Eq. (11)). Shaded bands indicate the uncertainty from a variation in the segment size of ACTR by ± 0.1 nm.

**Depletion interactions stabilize the IDP complex**

Fig. 2*C* shows that the complex between ACTR and NCBD is increasingly stabilized both with increasing crowder concentration and crowder size. From $K_D = k_{off} / k_{on}$ (which yields, within error, the same $K_D$ as calculated from the transfer efficiency histograms, see SI Table 1), we obtain the effect of crowding on the free energy of binding between the two IDPs according to[53]:

$$\Delta\Delta G = \Delta G - \Delta G_0 = -k_B T \ln \frac{K_{D,0}}{K_D} \ . \tag{2}$$

$\Delta G$ and $\Delta G_0$ are the binding free energies in the presence and absence of crowder, respectively; $K_{D,0}$ is the equilibrium dissociation constant in the absence of crowder ($K_{D,0}$ = 25±2 nM); $k_B$ and $T$ are the Boltzmann constant and absolute temperature, respectively. The largest measured stabilization by about an order of magnitude in $K_D$, or $\Delta\Delta G \approx$ -2 $k_B T$, was observed in 0.13-0.17 g/ml of the largest PEGs (4600-35000).[‡] What is the cause of this crowder size- and concentration-dependent stabilization of the ACTR-NCBD complex?

A commonly employed framework for crowding effects is scaled-particle theory[55], which estimates the free energy required for creating a cavity of the size of the biomolecules of interest in a solution of hard spheres equivalent to the size of the crowder. The total volume occupied by the two individual IDPs is greater than that of their folded complex, so complex stabilization with increasing crowder *concentration* is expected because the solution volume available to the proteins decreases[45,56]. However, scaled-particle theory predicts that with increasing crowder *size* (at fixed volume (or mass) fraction of crowding agent), the free energy cost for creating a cavity of a given volume decreases[57], and so complex stabilization should decrease, the opposite of what we observe (Fig. 2*C*). This marked discrepancy is reminiscent of the effect of polymeric crowders on the chain dimensions of IDPs[19] and indicates that a different theoretical framework is required.

Here, we utilize the concept of depletion interactions[38-40], which allows us to combine the influence of polymer physics[19] with the attractive interactions between particles (the proteins in our case) in a solution of crowders[40], as well as to describe the effect of crowders on viscosity and association kinetics. The origin of these effects is the existence of a depletion layer around a colloidal particle with radius *R*, in a solution of polymeric crowders with radius of gyration $R_g$ (Figs. 2A, S1*D*)[40]. The segments of the polymer cannot penetrate the particle, which leads to a loss of configurational entropy of the polymer near the surface of the colloid and thus a vanishing concentration of polymer segments in a layer around the surface. The thickness, *δ*, of this depletion layer is proportional to $R_g$ of the polymer in the dilute regime, whereas it depends on the average mesh size, $\xi$, in the semidilute regime (Fig. S1*E*)[58]:

$$\delta^{-2} = \delta_0^{-2} + \xi^{-2}, \tag{3}$$

---

[‡] For the smallest crowders (up to PEG 200), the stabilizing trend reverts at a number density of PEG above ~1/nm³ (Fig. 2*D*), possibly caused by the repulsive interactions between two particles at high concentrations of small crowders, as observed in optical tweezer experiments[54], owing to entropically stabilized layers of small crowders filling the inter-particle space. Since such contributions go beyond the excluded-volume effects of interest here, we restrict our analysis to data points with *n* < 1 nm⁻³. Another effect we do not consider here is the stabilization of the complex by ethylene glycol (and to a lesser extent by di- and triethylene glycol), which is not caused by excluded-volume effects but by the known unfavorable chemical interactions of the terminal hydroxyl groups with proteins[47].

where $\delta_0$ is the thickness of the depletion layer in dilute solution. A common approach to quantify the resulting attractive depletion force is *via* the osmotic pressure in a solution of polymers, $\Pi = n k_B T$, where $n$ is the number density[§] of polymer. If the particles are far apart, they are uniformly surrounded by polymers, and the resulting osmotic pressure around them is isotropic. If instead the depletion layers of the particles overlap, polymer chains cannot enter between them, leading to a non-isotropic osmotic pressure that pushes the particles together. Their distance-dependent attractive interaction potential, $W(r)$, then results as the product of the overlap volume of the depletion layers, $V_{\text{overlap}}(r)$, and $\Pi$: $W(r) = -n k_B T V_{\text{overlap}}(r)$. We assume that the net stabilization of the complex, $\Delta\Delta G$, corresponds to the interaction potential of the two particles at contact, $W(0)$ (which has previously been suggested to be a reasonable approximation for proteins[59]):

$$\Delta\Delta G = W(0) = -n k_B T V_{\text{overlap}}(0) = -\frac{c}{M} k_B T V_{\text{overlap}}(0), \qquad (4)$$

where *c* is the mass concentration of polymeric crowder. Eq. (8) (Theory) describes the calculation of $V_{\text{overlap}}(0)$ for two interacting spherical particles. Since larger crowders increase the size of the depletion layer and thus $V_{\text{overlap}}$, Eq. (4) rationalizes the observed stabilization of the ACTR-NCBD complex both with increasing crowder concentration and increasing crowder size. But can this simple theory account for our experimentally observed extent of stabilization quantitatively?

The dependence of $\Delta\Delta G$ on crowder concentration is approximately linear (Fig. 2E), as predicted by Eq. (4). However, the magnitude of this concentration dependence for two particles with radii corresponding to the experimentally determined hydrodynamic radii of ACTR and NCBD, clearly overestimates the experimentally observed stabilization (black dashed line in Fig. 2*F*, for details of the calculation, see Theory). Within the framework of Eq. 4, this discrepancy indicates that $V_{\text{overlap}}$ for the two proteins is too large. Fig. 2*B* illustrates that the size of the relevant depletion layer around an IDP is indeed expected to be much smaller than around a globular protein, because the polymeric crowders can penetrate the hydrodynamic sphere of the IDP, which is a polymer itself. The blue line in Fig. 2*F* shows the prediction of the stabilization calculated for two particles in dilute crowder solution, one with $R_1$ = 1.74 nm (corresponding to $R_H$ of NCBD, which is rather compact owing to its molten-globule-like character), and one with $R_2$ = 0.4 ± 0.1 nm, corresponding to the approximate size of a chain segment of an IDP such as ACTR. We note that $R_2$ is the only adjustable parameter in this framework. The resulting small stabilization of the complex describes the data up to PEG 2050 reasonably well[**], but above, it predicts an effective destabilization, in contrast to the experimental observation. For larger PEGs, however, we leave the dilute regime already at low PEG concentrations, so the overlap volume becomes a function of the correlation length[60], and the osmotic pressure must be treated in terms of blobs of volume $\sim \xi^3$ and concentration $\sim \xi^{-3}$ (since PEG fills the solution completely at $c > c^*$). We thus use a corresponding expression from renormalization group theory for the osmotic pressure in the semidilute regime[60] (see Theory), with $R_1$ = 1.74 nm and $R_2$ = 0.4 ± 0.1 nm. The result indeed agrees with the experimentally observed stabilization of the protein complex reasonably well, even for large crowder sizes (orange line in Fig. 2*F*).

---

[§] We treat number density (or number concentration) as equivalent to molar concentration. According to the 2019 redefinition of the SI units, 1 mol ≡ 6.02214076·10²³ particles, so any equation in terms of number density can be used equivalently in terms of molar concentrations simply by multiplying with 1 = 6.02214076·10²³/1 mol.

[**] The failure to capture the interaction free energy in ethylene glycol might arise from additional interactions of the ethylene glycol monomer with the proteins[33].

The pronounced improvement compared to the simple picture of the interaction between two spherical colloidal particles in dilute crowder solution suggests that the polymeric properties of both the crowders and the IDPs need to be taken into account: for the IDPs in terms of the relevant overlap volume of the highly disordered ACTR; for the polymeric crowders in terms of a decrease of the correlation length in the semidilute regime, where the chains overlap and screen each other's excluded-volume interactions. Notably, $\Delta\Delta G$ approaches saturation for PEGs with M $\gtrsim$ 4600 g/mol, as expected for polymers above $c^*$, where $\xi$ is independent of $P$.[40,52]

**Diffusion in a solution of polymeric crowders**

A key contribution to the rate of binding is the diffusivity of the interaction partners. Since in our measurements ACTR is surface-immobilized, we only need to account for the diffusivity of NCBD in solutions with different sizes and concentrations of PEG. We obtain the translational diffusion coefficients from FCS measurements of the acceptor-labeled NCBD in the solution directly above the surface by measuring the diffusion time, $\tau_D$, through the confocal volume and relating it to the diffusion time in the absence of crowder (Figs. 3, S3*A*,*B*). The diffusion coefficient of NCBD without crowder follows from its $R_H$ and the Stokes-Einstein relation as $D_0 = 1.3 \cdot 10^8$ nm$^2$s$^{-1}$. The diffusion coefficients in the presence of crowders, $D_1$, result from the corresponding measured diffusion times (Fig. S3*D*) as $D_1 = D_0 \tau_{D,0} / \tau_D$, with index '0' specifying the value in the absence of crowder. According to the Stokes-Einstein equation, $1/D_1$ is expected to scale with the bulk viscosity of the solution, $\eta_{\text{bulk}}$, as $1/D_1 \propto 6\pi\eta_{\text{bulk}}R_H / k_B T$ (Figs. 3*B*, S3*E*), where $R_H$ is the hydrodynamic radius of the diffusing particle. Up to PEG 2050 ($R_g$ = 1.8 nm), where the crowders are smaller than or similar in size to NCBD ($R_H$ = 1.74 nm), this relation describes the data reasonably well (Fig. 3*B*), but pronounced deviations are apparent for larger PEGs. In the presence of 0.1 g/l PEG 35000, e.g., the observed diffusion time of NCBD corresponds to only ~20% of the value expected for $\eta_{\text{bulk}}$. We quantify the observed microscopic viscosity relevant for the translational diffusion of NCBD, $\eta_{\text{micro}}$, according to $\eta_{\text{micro}} / \eta_s = \tau_D / \tau_{D,0} = D_0 / D_1$, where $\eta_s$ is the viscosity of the solution in the absence of crowders (1.0 mPa s at 22°C).

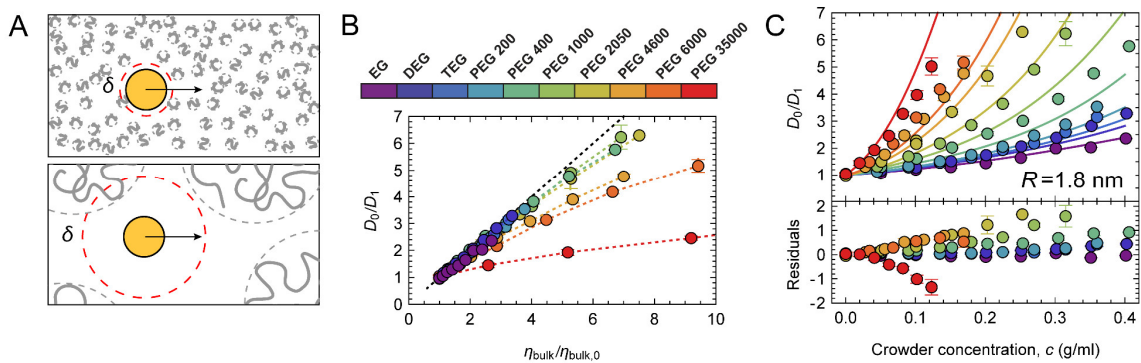

**Figure 3. Probing the microviscosity of PEG solutions with FCS.** (A) Diffusion of a particle through a solution of crowders with different sizes, with the respective thickness of the depletion layer, $\delta$ (red dashed circles). The larger the depletion layer, the less the diffusion of the particle is influenced by the bulk viscosity of the solution. (*B*) Relation between relative bulk viscosity and microviscosity, quantified via the translational diffusion coefficient of NCBD according to $\eta_{\text{micro}} / \eta_s = D_0 / D_1$ for solutions containing different concentrations and sizes of PEG (color scale above), where $\eta_{micro}$ is the microscopic viscosity, $\eta_s$ the neat solvent viscosity, $D_0$ is the diffusion coefficient in the absence, $D_1$ in the presence of crowder. The black dashed line shows the behavior expected from the Stokes-Einstein equation based on the bulk viscosity. (*C*) Microviscosity analyzed with a model using solvent viscosity within the depletion layer and bulk viscosity outside (Eq. (5)). A global fit of $D_0 / D_1 = \eta_{micro} / \eta_s$ across all PEG sizes and concentrations yields *R* = 1.8±0.1 nm for NCBD.

The theory of depletion interactions provides an adequate framework for describing the effect of microviscosity as probed by the diffusion of a molecule in a solution of polymeric crowders. The observed dependence on PEG size can be explained by the larger thickness of the depletion layer around NCBD in the presence of larger polymeric crowders (Fig. 3A). Within the depletion layer, where the polymer segment concentration is reduced, the microviscosity is expected to be closer to the viscosity of pure solvent. Hence, the larger the depletion layer around the particle, the less the particle is influenced by $\eta_{bulk}$. This effect can be described by the theory of Tuinier *et al.*[61], according to which

$$\eta_{micro} = \eta_s \frac{Q(\lambda, \varepsilon)}{Z(\lambda, \varepsilon)}, \quad (5)$$

where $Q(\lambda, \varepsilon)$ and $Z(\lambda, \varepsilon)$ are algebraic functions of the ratio of solvent and bulk viscosity, $\lambda = \eta_s / \eta_{bulk}$, and of the ratio of depletion layer thickness and particle radius, $\varepsilon = \delta / R$ (see Theory). The known values of $\delta$, $\eta_s$, and $\eta_{bulk}$ are used in a global fit of the diffusion data for all PEG sizes and concentrations (Fig. 3*C*). The single free fit parameter is *R*, which yields a value of 1.8 ± 0.1 nm, remarkably close to the size of NCBD[42] ($R_H$ = 1.74 nm). The fit is best for small PEGs, but even for larger PEGs, the theory predicts the observed microviscosities to within ~25%, suggesting that depletion effects are the dominant contribution to the low microviscosity experienced by NCBD in the presence of large PEGs.

**Depletion effects influence the association rate**

The kinetics of binding under crowded conditions are expected to be influenced by both of the depletion effects discussed above[62]. On the one hand, the crowder-induced attractive interaction potential should accelerate binding; on the other hand, the reduced diffusion coefficient should decelerate it. Based on the quantitative analysis of these two competing effects in the previous sections, we can now analyze their joint influence. An expression recently derived by Berezhkovskii and Szabo explicitly combines the two effects on the association rate coefficient, $k_{on}$ (Fig. 4*A*):[63]

$$\frac{1}{k_{on}} = \left( \frac{1}{k_0} + \frac{1}{4\pi D_0} \left( \frac{1}{R_{contact}} - \frac{1}{R_{cavity}} \right) \right) e^{-\frac{\Delta \Delta G}{k_B T}} + \frac{1}{4\pi D_1 R_{cavity}} \quad (6)$$

This special case of the Collins-Kimball-Debye[64,65] formula generalized to a distance-dependent diffusivity[66] accounts for the following effects:

(i) In the crowded solution, the reactants diffuse relative to each other with a diffusion coefficient $D_1$. Crowding decreases $D_1$ with respect to $D_0$, their relative diffusivity in pure solvent ($D_1 < D_0$, cf. Fig. S3*D*), which slows down association. Since ACTR is surface-immobilized, only the diffusion coefficient of NCBD needs to be considered in our case, which was quantified in the previous section (Fig. 3).

(ii) Once the reactants come within their contact radius, $R_{contact}$, they form a product with the intrinsic/reaction-controlled rate constant, $k_0$.

(iii) If the crowders are sufficiently large, they can accommodate the reactants within a cavity of radius $R_{cavity}$ that is devoid of crowders, so the diffusion coefficient of the reactants within the cavity is $D_0$. This effect speeds up the association reaction if $R_{cavity} > R_{contact}$, since the proteins can make contact faster than if they are separated by the crowder solution. $R_{cavity}$ is related to the thickness of the depletion layer, $\delta$, around the proteins. We calculated $\delta$ using Eq. (3) for a sphere with $R$ = 1.74 nm, the size of NCBD, and introduce a proportionality factor, $a$, yielding $R_{cavity} = a \cdot \delta$. We assume $a > 1$, i.e., $R_{cavity} > \delta$, since the cavity needs to accommodate two proteins.

(iv) A square-well potential localized in the cavity devoid of crowders leads to an attraction between the reactants and increases the association rate coefficient. As previously suggested,[67] we assume that the depth of the potential equals the depletion interaction free energy, $\Delta\Delta G$, which we measured as a function of crowder concentration (Fig. 2).

To probe the competing effects of viscosity-induced deceleration and depletion-induced acceleration of binding, we extracted $k_{on}$ from the single-molecule time traces (Fig. 1) recorded over the entire range of PEG sizes and concentrations (Fig. 4*B*). Overall, $k_{on}$ tends to exhibit an initial increase at low crowder concentrations, which is most pronounced for the largest PEGs. At higher crowder concentrations, the trend is reversed, and association slows down again – exactly the nonmonotonic behavior predicted by the competing effects that contribute to Eq. (6): The initial acceleration is caused by the attractive potential between the reactants induced by the depletion force, whereas at higher crowder concentrations, the strong decrease of $D_1$ due to the increase in viscosity dominates and leads to a deceleration.[62,63]

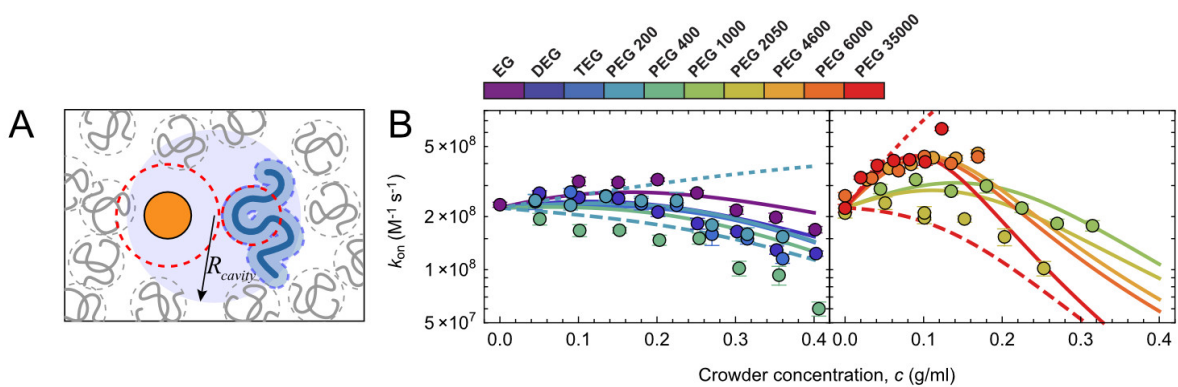

**Figure 4. Depletion effects on the association rate coefficient.** (*A*) Schematic depiction of NCBD (orange) and ACTR (blue) within a cavity formed by the crowder molecules. Inside the cavity, the proteins diffuse with the diffusion coefficient in pure solvent, outside with a reduced relative diffusion coefficient. Within the cavity, the proteins experience an attractive interaction due to the osmotic pressure of the crowders. (B) Global fit of the association rate coefficients, $k_{on}$, (filled circles) with Eq. (6) (lines) as a function of the concentration of PEGs of different size (see color scale). For clarity, the dataset is split in two: the *left panel* shows PEGs up to 400 g/mol, the *right panel* 1000 g/mol and above. The counteracting effects of depletion attraction (short-dashed lines) and viscosity (long-dashed lines) predicted by Eq. (6) are illustrated for PEG 200 (left) and PEG 35000 (right).

We fit all data in Fig. 4*B* globally using Eq. (6), with $R_{\text{contact}}$, $k_0$, and $a$ as shared fit parameters[††]. The resulting fit yields $R_{\text{contact}}$ = 0.54 ± 0.06 nm, $k_0$ = (4.0 ± 0.5)·10$^8$ M$^{-1}$ s$^{-1}$, and $a$ = 1.8 ± 0.1, and captures the overall behavior. The small value of $R_{\text{contact}}$ indicates a relatively compact encounter complex, in line with recent measurements[44], and $k_0$ is only about a factor of two lower than the purely diffusion-limited rate constant, $4\pi D_0 R_{\text{contact}}$, in keeping with the low association barrier of the protein pair identified previously[44]. The value of *a* suggests that the cavity radius is roughly twice the thickness of the depletion layer, which appears reasonable. A turnover and subsequent drop in $k_{\text{on}}$ is also predicted for the largest PEGs (4600-35000 g/mol), but only at crowder concentrations that were experimentally inaccessible owing to increased fluorescence background at high PEG concentrations. Finally, given the quantitative description of both the stability (Fig. 2) and the association kinetics of ACTR-NCBD binding (Fig. 4) based on depletion interactions, we note that the dissociation kinetics can be inferred according to $k_{\text{off}} = K_D k_{\text{on}}$.

## Discussion

Based on a comprehensive data set that covers a broad range of crowder sizes and reports on the effects of (macro)molecular crowding on microviscosity and the equilibrium and kinetic properties of a coupled folding and binding reaction, we present an integrated analysis based on depletion interactions[38-40]. The concept of depletion interactions enables us to combine the influence of polymer physics, such as chain overlap and excluded-volume screening[19], with the attractive interactions between proteins caused by the crowding agent (or depletant)[40], as well as with the effect of crowders on microviscosity and association kinetics[63]. In this way, the transition from the dilute to the semidilute regime can be treated quantitatively, which has previously been shown to be essential for understanding protein-protein interactions[33] and IDP dimensions[19] under the influence of polymeric crowding agents[‡‡]. The approach thus goes beyond the more commonly employed scaled particle theory[55], which is based on the free energy of insertion of a particle into a hard-sphere fluid and successfully describes many crowding-induced phenomena[45,56,57], especially as a function of crowder concentration. However, at fixed volume or mass fraction of crowder, scaled particle theory predicts for a process such as coupled folding and binding that the stability of the complex decreases with increasing crowder size, because the free energy cost for creating a cavity is smaller for larger crowders[19,55]. This trend is opposite to what we observe here experimentally (Fig. 2), illustrating the need for extending the theoretical approach.

We find that all our observations can be explained by depletion effects of the polymeric crowders on the interacting proteins. The decrease in polymer segment concentration near a colloidal particle (the proteins in our case) creates a cavity within which the protein diffuses according to the viscosity of the crowder-free solvent rather than the bulk viscosity. This effect on translational diffusion

---

[††] For the smallest crowder, ethylene glycol, $R_{\text{cavity}} < R_{\text{contact}}$, which would lead to unphysical results in Eq. (6). We thus set $1/R_{\text{contact}} - 1/R_{\text{cavity}} = 0$ in this case, for which Eq. (6) reduces to

$$1/k_{\text{on}} = 1/k_0\, e^{-\frac{\Delta\Delta G}{k_B T}} + 1/4\pi D_1 R_{\text{contact}}.$$

[‡‡] We note that we neglect the effect of crowding on IDP dimensions in the analysis of coupled folding and binding. As shown by Soranno *et al.*[19], IDPs that are already quite compact, such as the molten-globule-like NCBD[7,42,44], exhibit negligible compaction upon crowding with PEG. For ACTR, chain compaction by up to ~10% is expected at high crowder concentrations and sizes[19,50]. However, even this contribution is negligible compared to the decrease in relevant segment size to 0.4 nm we need to invoke for explaining the stability of the complex (Fig. 2).

is only expected if the size of the cavity is larger than the protein, so that it can effectively slip through the polymer network. This length scale dependence has previously been observed by varying the diameter of the probe instead of the crowder:[68] When the crowder-related length scale was larger than the probe, translational diffusion was significantly faster than the bulk value. The theory by Tuinier *et al.*[61] accounts for these different length scales and successfully predicts the observed translational diffusion coefficient. Similarly, depletion-enhanced diffusion relative to the bulk viscosity is the basis for the acceleration of binding that we observe for large crowders, as explained by a recent model for the influence of crowding on bimolecular association rates[63]. Accounting for the acceleration of binding additionally requires an attractive potential[63], which is caused by the depletion interactions between the two proteins and can be described by the theory of Asakura and Oosawa[38] extended to the semidilute regime[58,60]. Finally, it is worth pointing out that we also need to consider the disordered nature of ACTR explicitly, since the relatively low equilibrium stabilization we observe suggests that the relevant length scale for the depletion interactions of ACTR corresponds to the size of a chain segment, implying that IDPs should be less affected by crowding than folded proteins. Indeed, it has been demonstrated that the translational diffusion of IDPs is less reduced by crowding than that of globular proteins of the same size[69].

The combined framework of depletion interactions and polymer physics may be useful for quantitatively describing the effect of crowding on a wide range of biopolymers. For instance, single-molecule measurements have been used to investigate the effect of crowding on hairpin formation in RNA[70] and ribozyme compaction[71]. The results indicate a stabilization of more compact structures upon addition of PEG 8000, similar to the effects observed here. The crowder concentrations we use also cover the range of cellular concentrations of macromolecules (~0.1-0.15 g/ml in eukaryotes, 0.2-0.4 g/ml in *E. coli*)[12-14,72,73]. It will be interesting to investigate the relevant crowding length scales directly in the cell[14,32,74] and compare how diffusion, binding equilibria, and kinetics compare with the depletion interactions we observed *in vitro*. Extensions of the concepts used here may be required to account for the additional contributions from both globular and polymeric crowders as well as meshworks and small solutes in the cell.

## Theory

**Relevant length scales in the dilute and semidilute regimes**

For our analysis, we used the values for the radius of gyration of PEG, $R_g$, according to its dependence on *P*, the degree of polymerization, $R_g = 0.21 \text{ nm} \cdot P^{0.583}$,[75] where the scaling exponent indicates that water is a good solvent for PEG[51]. Each of the polymer chains occupies on average a volume $V = (4\pi/3)R_g^3$. The scaling law for the correlation length in the good-solvent regime[52], $\xi \approx R_g (c/c^*)^{-0.77}$, indicates that $\xi$ decreases steeply with increasing polymer concentration. Equivalently, $\xi \propto b^{-1.335}(cN_A/M_{monomer})^{-0.77}$, where $b$ is the segment length of PEG, and $M_{monomer}$ is the molar mass of a monomer. This relation shows that in the semidilute regime, $\xi$ is independent of $P$ and only a function of the polymer concentration (which follows from substituting the length scaling of $R_g$ into Eq. (1))[40].

**Depletion interactions**

The basis of depletion interactions between particles of radius $R$ in a solution of polymers is that the segments of the polymer cannot penetrate the particle, which leads to a loss of configurational entropy of the polymer near the surface of the colloid and thus a vanishing concentration of polymer segments in a depletion layer around the surface. In the dilute regime, we calculate the change in interaction free energy, $\Delta\Delta G$, due to the depletion layer with the classic Asakura-Oosawa model[38], assuming that the net stabilization corresponds to the depletion potential at contact, $W(0)$:

$$\Delta\Delta G = W(0) = -n k_B T V_{overlap}(0) = -\frac{c}{M} k_B T V_{overlap}(0) \tag{7}$$

The terms $n k_B T$ or $\frac{c}{M} k_B T$, respectively, describe the osmotic pressure, $\Pi$, where $n$ is the number density (or molar concentration§), $c$ the mass concentration, and $M$ the molar mass of the polymeric crowder. The overlap volume, $V_{overlap}(0)$, of two spherical particles of radii $R_1$ and $R_2$ in contact, with depletion layers of thickness $\delta_s$, is calculated based on elementary geometrical considerations according to

$$V_{overlap} = \frac{\pi(r+R-d)^2 \left(d^2 - 3(r-R)^2 + 2d(r+R)\right)}{12d}, \tag{8}$$

with $r = R_1 + \delta_s(R_1)$, $R = R_2 + \delta_s(R_2)$, $d = R_1 + R_2$.

A common approximation for the depletion layer in a solution of polymers is to replace the resulting smooth segment concentration profile near the particle surface by a step function that is zero up to a depletion layer thickness, $\delta$, and equal to the bulk concentration above[76]. $\delta$ then corresponds to the thickness of the layer around the particle surface from which the centers of mass of the polymer chains are excluded (Fig. S1$D$). In the dilute regime, $\delta \propto R_g$ for $R \gg R_g$ – the smaller the crowder, the closer its center of mass can be to the colloidal particle. The depletion layer thickness near a flat plate in a dilute solution of excluded volume polymers with radius of gyration $R_g$ was calculated by Hanke et al. using renormalization group theory (RGT)[40,77]:

$$\delta_0 = 1.07 R_g \tag{9}$$

The conversion of $\delta_0$ near a flat plate to the corresponding value, $\delta_s$, near a sphere with radius $R$ is a geometrical problem. If $R_g$ is similar to or greater than $R$, a correction term needs to be introduced to account for the interpenetration between particle and polymers. For excluded volume chains in the dilute regime (up to PEG 2050), we use the following expression that has been found using RGT[60,77]:

$$\frac{\delta_s}{R} = \left[1 + 3\frac{\delta_0}{R} + 2.273\left(\frac{\delta_0}{R}\right)^2 - 0.0975\left(\frac{\delta_0}{R}\right)^3\right]^{1/3} - 1. \tag{10}$$

To calculate the depletion potential at contact in the semidilute regime (polymer concentration $c > c^*$), we employed a relation based on the generalized Gibbs adsorption equation:[60]

$$\Delta\Delta G = W(0) = -k_B T \int_0^n \frac{1}{n'}\left(\frac{\partial \Pi}{\partial n'}\right)\left(\Gamma(0,n') - \Gamma(\infty,n')\right) dn', \text{ with } n = \frac{c}{M}. \tag{11}$$

Eq. (11) is also valid in the dilute regime, where it simplifies to Eq. (7). The expression for the osmotic compressibility, $\partial \Pi / \partial n$, based on RGT[78], is:

$$\left(\frac{\partial \Pi}{\partial n}\right) = 1 + 2.63\phi \left(\frac{1 + 3.25\ \phi + 4.15\ \phi^2}{1 + 1.48\ \phi}\right)^{0.309}, \text{ with } \phi = \frac{c}{c^*}. \tag{12}$$

$\Gamma(h,n)$ corresponds to the (negative) amount of adsorbed polymer segments when the spheres are a distance $h$ apart. It equals the product of $n$ and the overlap volume, thus:

$$\Gamma(0,n) = nV_{overlap} \text{ and } \Gamma(\infty,n) = 0. \tag{13}$$

The overlap volume was again calculated with Eq. (8), but in this case with $\delta_s$ evaluated in the semidilute regime, where the size of the interacting entity (the 'blob') is determined by $\xi$, so $\delta$ becomes a function of $\xi$ instead of $R_g$, and $\delta \approx \xi$ in the semidilute regime.[52] To this end, we employed a simple relation derived by Fleer *et al.* for calculating the depletion thickness near a flat plate in the semidilute regime[58], which we used for all PEG sizes and concentrations:

$$\delta^{-2} = \delta_0^{-2} + \xi^{-2}, \tag{14}$$
$$\text{with } \xi = R_g (c/c^*)^{-0.77} \text{ and } c^* = 3M_p / (4\pi N_A R_g^3),$$

and used $\delta$ instead of $\delta_0$ in Eq. (10) for calculating the thickness of the depletion layer around a sphere in the semidilute regime.

**Diffusion through a solution of polymers**

Tuinier et al. derived an expression for calculating the friction coefficient, *f*, of a spherical particle as a function of the ratio of the thickness of the depletion layer to the size of the particle, $\varepsilon = \delta / R_H$:[61] $f = 6\pi\eta_s R_H Q(\lambda,\varepsilon) / Z(\lambda,\varepsilon)$, where $\lambda = \eta_s / \eta_{bulk}$. With the algebraic functions $Q(\lambda,\varepsilon)$ and $Z(\lambda,\varepsilon)$ (see *Theory*), and $\delta$ calculated as outlined above (Eq. (3)), the microviscosity is given by

The microviscosity experienced by a sphere with hydrodynamic radius $R_H$ that is diffusing through a polymer solution with bulk viscosity $\eta_{bulk}$ and solvent viscosity $\eta_{solvent}$ is calculated with the relation obtained by Tuinier et al.[61]

$$\frac{\eta_{micro}}{\eta_{micro,0}} = \frac{Q(\lambda,\varepsilon)}{Z(\lambda,\varepsilon)}, \text{ with}$$
$$Q(\lambda,\varepsilon) = 2(2+3\lambda)(1+\varepsilon)^6 - 4(1-\lambda)(1+\varepsilon) \text{ and}$$
$$Z(\lambda,\varepsilon) = 2(2+3\lambda)(1+\varepsilon)^6 - 9\left(1 - \frac{1}{3}\lambda - \frac{2}{3}\lambda^2\right)(1+\varepsilon)^5 \tag{15}$$
$$+ 10(1-\lambda)(1+\varepsilon)^3 - 9(1-\lambda)(1+\varepsilon) + 4(1-\lambda)^2$$
$$\varepsilon = \frac{\delta_s}{R}, \lambda = \frac{\eta_{solvent}}{\eta_{bulk}}$$

$\delta_s$, the depletion layer thickness around a sphere of radius *R*, was calculated from Eq. (10) with the approximation for the semidilute regime Eq. (14).

## Materials and Methods

**Protein expression, purification and labeling**

ACTR and NCBD were purified, expressed and labeled as described before.[7] Shortly, a single-cysteine Avi-tagged[79] ACTR variant was *in vivo*-biotinylated in *E.coli* and purified with immobilized metal ion chromatography (IMAC) via a C-terminal $His_6$-tag. The tag was cleaved off with thrombin, and the protein further purified with HPLC on a C18 column (Reprosil Gold 200, Dr. Maisch). Lyophilized protein was labeled with a 0.8fold molar ratio of Cy3B maleimide dye (GE Healthcare); the single-labeled protein was purified with HPLC (Sunfire C18, Waters).

A single-cysteine NCBD variant was co-expressed with ACTR as described before[41]. Purification was carried out using IMAC via an N-terminal $His_6$-tag. The tag was cleaved off with HRV 3C protease and the protein was further purified with HPLC on a C18 column (Reprosil Gold 200). Lyophilized protein was labeled with a 1.5fold molar ratio of CF680R maleimide dye (Biotium); the single-labeled protein was purified with HPLC (Reprosil Gold 200).

**Preparation of PEG solutions**

PEG solutions were prepared as described before.[19] Briefly, crowding experiments were carried out in 50 mM sodium phosphate buffer, pH 7.0 (NaP buffer). PEG solutions were prepared by mixing acidic (50 mM $NaH_2PO_4$ + PEG) and alkaline (50 mM $Na_2HPO_4$ + PEG) stock solutions to a final pH of 7.0 (+/- 0.05). PEG solutions of lower concentrations were prepared by diluting the corresponding stock solution to the desired concentration with NaP buffer. The pure (poly)ethylene glycols were purchased at Sigma, with exception of ethylene glycol (SPECTRANAL, Riedel-de Haën) and PEG 400 (ROTIPURAN, Roth). The bulk viscosity of the PEG solutions was measured with a digital rotational viscometer (DV-I+, Brookfield).

**Surface immobilization**

The single-molecule binding experiment was conducted as described before.[7] In short, adhesive silicone hybridization chambers (Secure Seal Hybridization Chambers, SA8R-2.5, Grace Bio-Labs) were fixed onto PEGylated, biotinylated quartz coverslips (Bio_01, MicroSurfaces, Inc.). 0.2 mg/ml Avidin D (Vector Labs) were incubated for 5 min in a reaction chamber, followed by addition of 10 pM ACTR-Cy3B to yield a surface coverage of 0.1-0.3 molecules/$\mu m^2$. Binding experiments were conducted in the appropriate PEG solution, supplied with 16 nM CF680R-labeled NCBD, 0.01% Tween 20, 1% (w/v) glucose, 0.4 mg/ml glucose oxidase, 400 U/ml catalase as oxygen scavenging system, as well as 1 mM methyl viologen and 1 mM ascorbic acids as triplet quenchers. Addition of these components led to a 0.1fold dilution of the PEG solution, which was taken into account.

**Single-molecule experiments**

All single-molecule experiments were conducted at 22°C on a MicroTime 200 (PicoQuant, Germany) equipped with a 532-nm cw laser (LaserBoxx LBX-532-50-COL-PP, Oxxius) and a 635-nm diode laser (LDH-D-C-635M, PicoQuant). Florescence photons were separated from the scattered laser light with a triple-band mirror (zt405/530/630rpc, Chroma). A dichroic mirror was used to separate donor and acceptor emission (T635LPXR, Chroma). Donor photons were filtered with an ET585/65m bandpass filter (Chroma), acceptor photons with a LP647RU long pass filter (Chroma), followed by detection with two SPCM-AQRH-14 single-photon avalanche diodes (Perkin Elmer). For FCS measurements, acceptor photons were split according to their polarization, filtered with LP647RU long-pass filters and detected on two SPCM-AQRH-14 single-photon avalanche diodes. To enable surface scanning, the objective (UPlanApo 60×/1.20-W, Olympus) was mounted on a piezo stage (P-733.2 and PIFOC, Physik Instrumente GmbH).

Single Cy3B-labeled ACTR molecules were localized on the surface as described before[7] and recorded at a laser power of 0.5 µW (measured at the back aperture of the objective) until photobleaching occurred. Time traces from a total of 30-40 molecules were recorded for each PEG concentration. Note that the fluorescence signal was not corrected for background, quantum yields, channel crosstalk etc. since none of the observables used for our analysis depends on these corrections. We thus only report apparent transfer efficiencies in Fig. 1*C*. Before and after recording the time traces at a given set of conditions, the diffusion time and concentration of NCBD was estimated from FCS measurements using the 635-nm diode laser (10 µW, measured at the back aperture of the objective). For this purpose, the laser was focused 20 µm above the cover slide surface where ACTR was immobilized, and the fluorescence signals of the two acceptor detection channels were cross-correlated.

**Analysis of single-molecule time traces**

Single-molecule time traces were analyzed as described before.[7] We previously showed that NCBD binds to ACTR in two conformations, NCBD1 and NCBD2, which correspond to the peptidyl-prolyl bond involving Pro20 being in trans or cis configuration, respectively[7]. The binding and dissociation rates of these states to ACTR differ, but the relative donor and acceptor photon rates of the bound states are identical. We used the following rate matrix to describe the kinetics of freely diffusing NCBD1 and NCBD2 interacting with a surface-immobilized ACTR molecule (see kinetic scheme in Fig. 2*E* of Zosel et al.[7]):

$$\mathbf{K}_{3state,blink} = \begin{pmatrix} -(k'_{on,1} + k'_{on,2} + k_{+b}) & k_{off,1} & k_{off,2} & k_{-b} \\ k'_{on,1} & -k_{off,1} & 0 & 0 \\ k'_{on,2} & 0 & -k_{off,2} & 0 \\ k_{+b} & 0 & 0 & -k_{-b} \end{pmatrix} \quad (16)$$

The first three states correspond to free ACTR, NCBD1 bound to ACTR, and NCBD2 bound to ACTR, respectively. The two association rates are given by $k'_{on,i} = k_{on,i} \cdot c_{NCBDi}$ with $i$ = 1,2. $k_{off,i}$ are the corresponding dissociation rate coefficients. An additional dark state was introduced to represent photon blinking of the donor dye while no NCBD is bound to ACTR (the fluorescence blinking occurring in other states can be neglected[7]). Assuming that the relative populations of NCBD1 and NCBD2 do not depend on the crowder concentration, we set $c_{NCBD2} = 0.56 c_{NCBD1}$ and determined $k_{on,1}$, $k_{on,2}$, $k_{off,1}$, and $k_{off,2}$ for all concentrations of crowders using the maximum-likelihood procedure based on a hidden Markov model as described previously[7]. The results, displayed in Fig. S2 and TableS1, show that the relative effects of crowders are identical for ACTR interacting with NCBD1 and NCBD2. Hence, we present in the main text a simpler analysis, in which we neglect the difference between the two binding kinetics for the sake of clarity. We thus analysed the data using the rate matrix

$$\mathbf{K}_{2state,blink} = \begin{pmatrix} -(k'_{on} + k_{+b}) & k_{off} & k_{-b} \\ k'_{on} & -k_{off} & 0 \\ k_{+b} & 0 & -k_{-b} \end{pmatrix}, \quad (17)$$

where $k'_{on} = k_{on} \cdot c_{NCBD}$ and $k_{off}$ are the observed association and dissociation rates. This procedure is justified because the relative dependencies of $k_{on}$ and $k_{off}$ on crowder size and crowder concentration are within error the same as for the individual $k_{on,i}$ and $k_{off,i}$ values. For both the 2-state and the 3-state analysis, the time traces were binned in 1-ms intervals. The error bars in Figs. 2, 4 and S2 were obtained from bootstrapping; ten synthetic data sets of photon traces were randomly sampled from the measured data and analyzed in the same way as the original data set. Error bars for all derived quantities were propagated.

**Fluorescence correlation spectroscopy**

The mean diffusion time, $\tau_D$, of NCBD molecules (labeled with CF680R) through the confocal volume and the average number of NCBD molecules in the confocal volume, $\langle N \rangle$, were determined before and after recording the time traces as described by Zosel et al.[7] (with the aspect ratio of the confocal volume set to 0.165). The variations between the two FCS measurements (before and after recording single-molecule time traces) are depicted as error bars in Fig. S3*C,D*. The values of $\langle N \rangle$ from all measurements (with exception of the highest concentrations of PEG 2050, 6000 and 35000, see Fig. S3*C*) were averaged to calculate the mean number of molecules, $\langle N \rangle_{\text{avg}}$, present in the confocal volume at an NCBD concentration of 16 nM. To account for preparative sample-to-sample variation, the NCBD concentrations were corrected with $c = 16\,\text{nM}\,\langle N \rangle / \langle N \rangle_{\text{avg}}$. The association rate coefficients were then calculated from $k_{on} = k'_{on} / c$.

**Acknowledgements**

We thank Karin Buholzer, Iwo König, Attila Szabo, and Huan-Xiang Zhou for helpful discussion. This work was supported by the Swiss National Science Foundation.

**Author contributions**

F.Z., A.S., and B.S. designed research; F.Z., A.S., and D.N. performed research; D.N. contributed analytical tools; F.Z., A.S., and D.N. analyzed data; F.Z. and B.S. wrote the paper with contributions from all authors.

# Supplementary Information

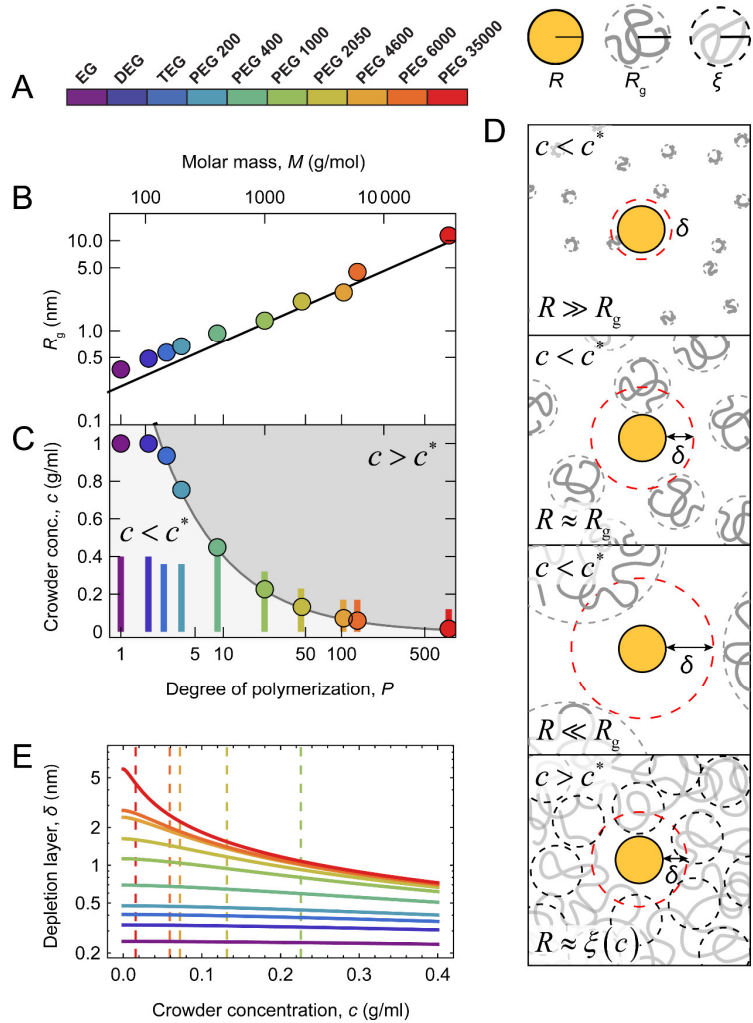

**Figure S1. Polymeric crowders and the depletion layer.** (*A*) Color legend for PEGs with different molar masses. (*B*) Radius of gyration, $R_g$, plotted against the degree of polymerization, $P$, and the molar mass, $M$, of PEG[80,81], and a fit with the scaling law $R_g = 0.21 \text{ nm} \cdot P^{0.583}$.[75] Deviations from the fit (which was obtained for PEG molecules over the entire range of lengths originally reported[75]) are due to finite-length effects for small values of $P$. (*C*) *Line*: overlap concentration, $c^*$, calculated from Eq. (1) with $R_g = 0.21 \text{ nm} \cdot P^{0.583}$, Circles: $c^*$ calculated based on published $R_g$ values[75,80,81]. *Bars*: Ranges of PEG concentrations probed here. (*D*) The depletion layer around a colloidal particle (modeled as hard spheres of radius *R*, orange) in solution with polymers of radius of gyration $R_g$ (gray). The thickness of the depletion layer, $\delta$, is visualized for several regimes: For $R \gg R_g$ (top panel), the depletion layer is given by Eq. (9) and increases with crowder size (second and third panels); for $R \approx R_g$, Eq. (10) is valid. In the semidilute regime ($c > c^*$, fourth panel), the polymer chains overlap and form a network with a mesh size given by the correlation length $\xi = R_g (c/c^*)^{-0.77}$, resulting in $\delta = \xi$. (*E*) $\delta$ as a function of the PEG concentration, for every size of PEG used, with the respective overlap concentrations indicated as dashed lines.

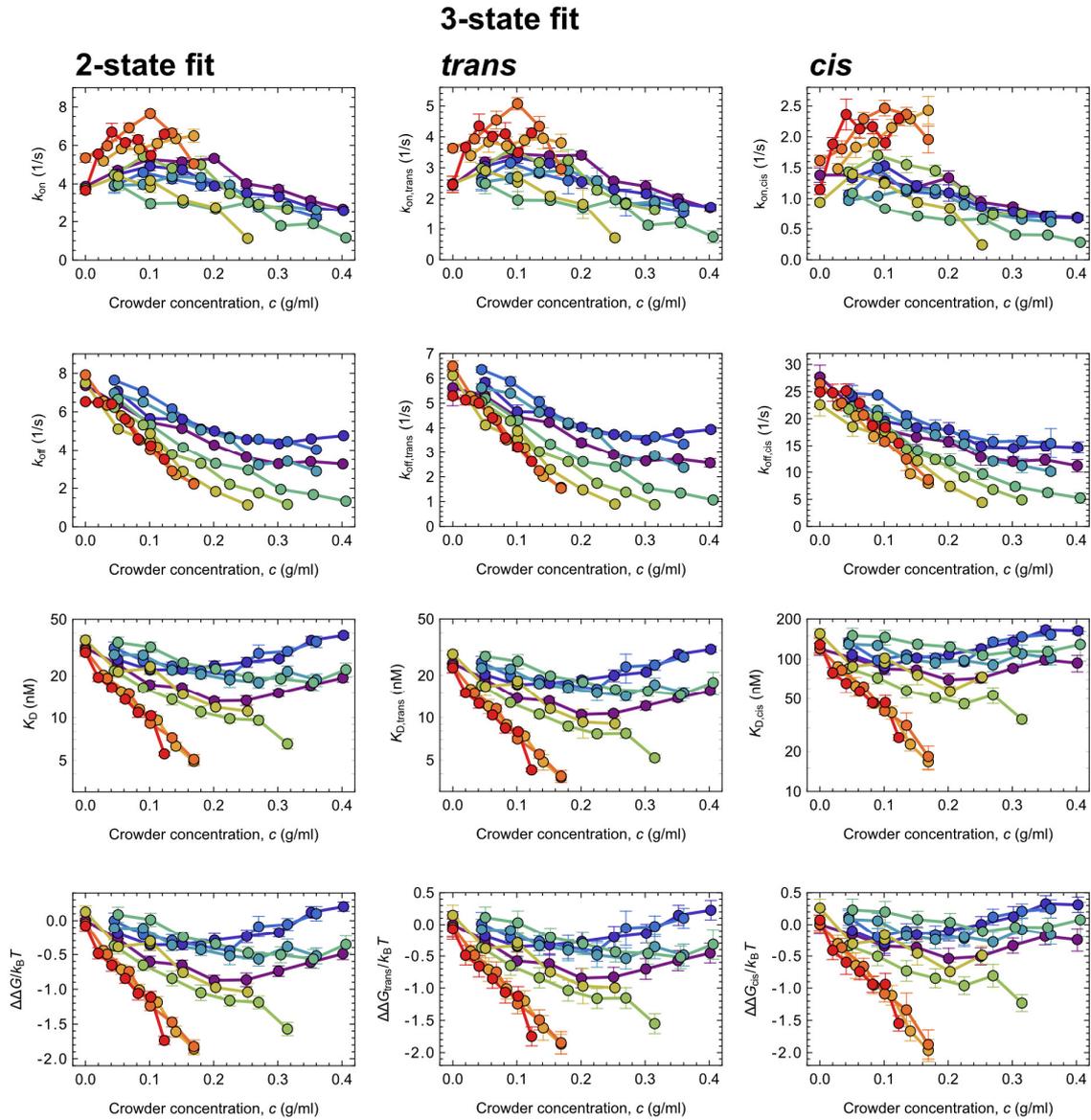

**Figure S2.** Comparison of kinetic and equilibrium parameters obtained from maximum likelihood analysis using either a two-state model (Eq. 17, left column) or a three-state model (Eq. 16, middle and right columns). The three-state analysis assumes that NCBD can exist in two conformations that differ in the configuration of a proline residue (trans or cis)[44]. In the two-state analysis (which we focus on in the main text for the sake of clarity), the two conformations are treated as equivalent in terms of their response to crowding, which is supported by this comparison. See Methods for details.

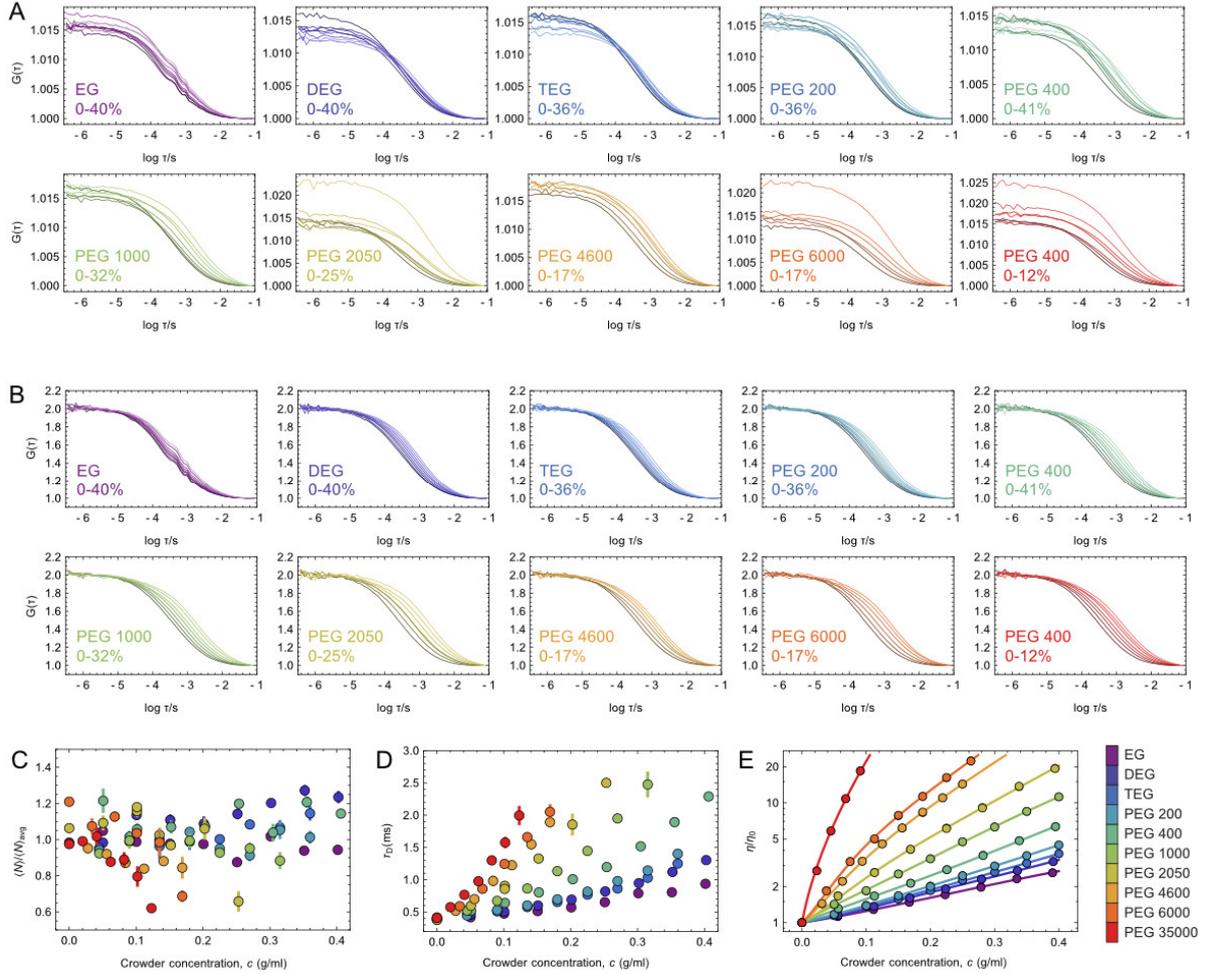

**Figure S3.** FCS measurements of freely diffusing NCBD. (*A*) Raw data. Two measurements were done per PEG concentration (one before and one after recording surface trajectories); their respective average is displayed. Lighter colors indicate increasing PEG concentrations, with the corresponding range indicated in each plot. All FCS curves were fitted with a free amplitude, a diffusion term, and a triplet term (shared among all fits, see Materials and Methods). The resulting values were used to quantify the change in viscosity as well as the concentration of NCBD. (*B*) FCS curves normalized to an amplitude of 1 to illustrate the increase in diffusion time with increasing PEG concentration. (*C*) Mean number of molecules in the confocal volume, $\langle N \rangle$, normalized by the average over all measurements, $\langle N \rangle_{\text{avg}}$ (*Materials and Methods*), plotted against the PEG concentration. Data points are the mean of two measurements, error bars indicate the span. For large PEGs (2050, 6000, 35000), there is a significant decrease in the number of NCBD molecules at high PEG concentrations. This effect is probably caused by adsorption of NCBD to the coverslip surface, as single molecule trajectories recorded under these conditions also display a significantly higher background in the acceptor channel. (*D*) Diffusion time of NCBD, plotted against the PEG concentration. Data points are the mean of two measurements, error bars indicate the span. (*E*) Bulk viscosities as a function of the PEG concentration, as measured by a shear-flow viscometer. The lines are polynomial interpolations of the data that were used as reference for fitting microviscosities.

**Table S1.** Experimental parameters recorded for the PEG titration series. *1st column*: Concentration of crowder. *2nd column*: total recording time of the experiment. *3rd column*: concentration of NCBD determined as the mean of two FCS measurements (error reflects the span of the measurements). *4th column*: viscosity relative to the viscosity in phosphate buffer ($\eta/\eta_0$), determined from the mean diffusion time of NCBD in two FCS experiments (error reflects the span of the measurements). *5-6th column*: fraction bound, determined from transfer efficiency histograms (histo) or rate coefficients (*columns 7-9*). *7th-9th column*: association and dissociation rate coefficients, and equilibrium dissociation constant, determined from a fit with a two-state model. *10th-15th column*: association and dissociation rate coefficients, and equilibrium dissociation constants, determined from a fit with a three-state model. Errors are propagated from the associated parameters.

**50 mM sodium phosphate buffer**

| $C$ (g/ml) | total time (s) | $c_{NCBD}$ (nM) | $\eta/\eta_0$ | frac. bound (histo) | frac. bound (rates) | $k_{on}$ (s$^{-1}$) | $k_{off}$ (s$^{-1}$) | $K_D$ (nM) | $k_{on,1}$ (s$^{-1}$) | $k_{on,2}$ (s$^{-1}$) | $k_{off,1}$ (s$^{-1}$) | $k_{off,2}$ (s$^{-1}$) | $K_{D,1}$ (nM) | $K_{D,2}$ (nM) |
|---|---|---|---|---|---|---|---|---|---|---|---|---|---|---|
| 0 | 2033 | 16.3±0.2 | 0.97±0.03 | 0.33 | 0.34±0.02 | 3.8±0.1 | 7.4±0.6 | 31.4±2.6 | 2.4±0.2 | 1.4±0.1 | 5.6±0.7 | 27.7±2.2 | 24.1±3.5 | 118±13 |
| 0 | 2468 | 17.7±0.2 | 0.95±0.03 | 0.33 | 0.33±0.02 | 3.7±0.2 | 7.5±0.3 | 35.2±2.7 | 2.5±0.3 | 0.9±0.1 | 6.1±0.2 | 22.6±2.1 | 27.9±1.8 | 153±13 |
| 0 | 2600 | 20.1±0.1 | 1.05±0.05 | 0.40 | 0.40±0.02 | 5.3±0.2 | 7.9±0.2 | 29.8±1.2 | 3.6±0.1 | 1.6±0.1 | 6.5±0.2 | 26.5±1.7 | 23.0±0.7 | 118±11 |
| 0 | 3263 | 16.2±0.2 | 1.03±0.03 | 0.35 | 0.36±0.03 | 3.7±0.3 | 6.6±0.5 | 29.0±2.9 | 2.5±0.3 | 1.1±0.1 | 5.3±0.1 | 24.9±2.8 | 22.3±1.7 | 126±16 |

**Ethylene glycol**

| $C$ (g/ml) | total time (s) | $c_{NCBD}$ (nM) | $\eta/\eta_0$ | frac. bound (histo) | frac. bound (rates) | $k_{on}$ (s$^{-1}$) | $k_{off}$ (s$^{-1}$) | $K_D$ (nM) | $k_{on,1}$ (s$^{-1}$) | $k_{on,2}$ (s$^{-1}$) | $k_{off,1}$ (s$^{-1}$) | $k_{off,2}$ (s$^{-1}$) | $K_{D,1}$ (nM) | $K_{D,2}$ (nM) |
|---|---|---|---|---|---|---|---|---|---|---|---|---|---|---|
| 0.05 | 2356 | 17.4±0.1 | 1.05±0.00 | 0.42 | 0.42±0.02 | 4.8±0.2 | 6.5±0.2 | 23.7±1.3 | 3.2±0.2 | 1.4±0.1 | 5.3±0.2 | 22.2±1.6 | 18.5±0.7 | 101±10 |
| 0.1 | 2563 | 16.4±0.2 | 1.20±0.03 | 0.49 | 0.49±0.02 | 5.3±0.2 | 5.5±0.1 | 17.0±0.7 | 3.4±0.1 | 1.3±0.1 | 4.4±0.1 | 18.1±1.2 | 13.5±0.4 | 83±8 |
| 0.151 | 2467 | 16.2±0.2 | 1.30±0.04 | 0.50 | 0.50±0.02 | 5.1±0.2 | 5.1±0.1 | 16.2±0.6 | 3.4±0.3 | 1.2±0.1 | 4.2±0.1 | 16.6±1.4 | 13.0±0.6 | 82±5 |
| 0.201 | 3245 | 16.2±0.1 | 1.44±0.03 | 0.55 | 0.55±0.03 | 5.3±0.2 | 4.3±0.2 | 13.0±0.8 | 3.4±0.1 | 1.3±0.1 | 3.4±0.1 | 15.8±1.0 | 10.3±0.4 | 69±8 |
| 0.251 | 2530 | 14.5±0.3 | 1.64±0.01 | 0.53 | 0.53±0.02 | 4.0±0.1 | 3.6±0.2 | 13.1±0.6 | 2.6±0.1 | 0.9±0.1 | 2.9±0.1 | 12.8±1.5 | 10.5±0.5 | 71±6 |
| 0.301 | 3253 | 16.9±0.1 | 1.99±0.00 | 0.53 | 0.53±0.04 | 3.7±0.2 | 3.3±0.2 | 14.9±1.2 | 2.4±0.2 | 0.9±0.0 | 2.6±0.0 | 12.0±0.6 | 11.9±0.8 | 84±4 |
| 0.351 | 3293 | 15.6±0.1 | 2.03±0.00 | 0.48 | 0.48±0.04 | 3.1±0.2 | 3.4±0.1 | 16.9±1.4 | 2.0±0.1 | 0.7±0.0 | 2.7±0.1 | 12.3±1.0 | 13.6±0.8 | 97±8 |
| 0.401 | 4283 | 15.7±0.3 | 2.36±0.00 | 0.45 | 0.45±0.04 | 2.7±0.2 | 3.3±0.3 | 19.0±2.3 | 1.7±0.1 | 0.7±0.0 | 2.6±0.2 | 11.2±1.1 | 15.3±0.7 | 92±13 |

**Diethylene glycol**

| $C$ (g/ml) | total time (s) | $c_{NCBD}$ (nM) | $\eta/\eta_0$ | frac. bound (histo) | frac. bound (rates) | $k_{on}$ (s$^{-1}$) | $k_{off}$ (s$^{-1}$) | $K_D$ (nM) | $k_{on,1}$ (s$^{-1}$) | $k_{on,2}$ (s$^{-1}$) | $k_{off,1}$ (s$^{-1}$) | $k_{off,2}$ (s$^{-1}$) | $K_{D,1}$ (nM) | $K_{D,2}$ (nM) |
|---|---|---|---|---|---|---|---|---|---|---|---|---|---|---|
| 0.05 | 2681 | 16.3±0.2 | 1.16±0.06 | 0.38 | 0.39±0.03 | 4.5±0.3 | 7.1±0.3 | 25.6±1.8 | 3.1±0.2 | 1.3±0.1 | 5.8±0.1 | 24.2±1.9 | 20.0±0.9 | 106±11 |
| 0.101 | 2568 | 18.8±0.3 | 1.28±0.02 | 0.46 | 0.46±0.03 | 4.9±0.2 | 5.6±0.3 | 21.7±1.6 | 3.3±0.2 | 1.5±0.1 | 4.7±0.3 | 19.6±0.8 | 16.9±0.6 | 87±9 |
| 0.151 | 3989 | 18.4±0.4 | 1.51±0.02 | 0.45 | 0.46±0.02 | 4.7±0.2 | 5.6±0.1 | 21.9±1.1 | 3.1±0.3 | 1.2±0.1 | 4.6±0.1 | 18.4±0.7 | 17.3±1.1 | 101±5 |
| 0.201 | 3841 | 18.1±0.5 | 1.74±0.00 | 0.43 | 0.44±0.03 | 3.9±0.2 | 5.0±0.5 | 23.2±2.6 | 2.5±0.2 | 1.1±0.1 | 4.0±0.1 | 18.0±1.3 | 18.3±1.0 | 107±14 |
| 0.252 | 4248 | 19.0±0.3 | 2.07±0.01 | 0.43 | 0.43±0.02 | 3.5±0.2 | 4.5±0.2 | 24.6±1.5 | 2.3±0.2 | 0.9±0.0 | 3.7±0.1 | 14.9±0.9 | 19.7±1.9 | 118±7 |
| 0.302 | 2883 | 20.0±0.1 | 2.39±0.01 | 0.43 | 0.43±0.01 | 3.3±0.1 | 4.3±0.1 | 26.0±0.7 | 2.2±0.1 | 0.8±0.0 | 3.5±0.1 | 14.5±1.6 | 20.8±0.9 | 133±16 |
| 0.352 | 3868 | 21.1±0.4 | 2.82±0.02 | 0.37 | 0.38±0.02 | 2.7±0.2 | 4.6±0.1 | 35.1±2.1 | 1.8±0.2 | 0.7±0.0 | 3.8±0.1 | 14.9±1.0 | 27.9±1.8 | 163±11 |
| 0.402 | 3033 | 20.5±0.4 | 3.28±0.03 | 0.35 | 0.35±0.02 | 2.5±0.1 | 4.7±0.1 | 38.1±2.1 | 1.7±0.1 | 0.7±0.0 | 3.9±0.1 | 14.7±1.1 | 30.2±1.5 | 161±9 |

**Triethylene glycol**

| $C$ (g/ml) | total time (s) | $c_{NCBD}$ (nM) | $\eta/\eta_0$ | frac. bound (histo) | frac. bound (rates) | $k_{on}$ (s$^{-1}$) | $k_{off}$ (s$^{-1}$) | $K_D$ (nM) | $k_{on,1}$ (s$^{-1}$) | $k_{on,2}$ (s$^{-1}$) | $k_{off,1}$ (s$^{-1}$) | $k_{off,2}$ (s$^{-1}$) | $K_{D,1}$ (nM) | $K_{D,2}$ (nM) |
|---|---|---|---|---|---|---|---|---|---|---|---|---|---|---|
| 0.045 | 4140 | 15.8±0.3 | 1.16±0.02 | 0.33 | 0.34±0.03 | 3.9±0.4 | 7.7±0.2 | 31.0±2.9 | 2.6±0.4 | 1.1±0.1 | 6.4±0.1 | 24.8±1.7 | 24.3±2.6 | 132±9 |
| 0.09 | 3308 | 16.4±0.3 | 1.29±0.02 | 0.39 | 0.39±0.03 | 4.5±0.3 | 7.0±0.4 | 25.3±2.1 | 3.2±0.2 | 1.5±0.1 | 5.9±0.2 | 24.4±0.4 | 19.6±1.2 | 97±8 |
| 0.135 | 4640 | 16.0±1.0 | 1.46±0.02 | 0.40 | 0.41±0.02 | 4.2±0.1 | 6.1±0.3 | 23.2±1.2 | 2.8±0.1 | 1.1±0.1 | 5.1±0.2 | 20.6±0.8 | 18.4±0.5 | 106±6 |
| 0.18 | 4414 | 16.4±0.6 | 1.65±0.02 | 0.43 | 0.43±0.04 | 3.9±0.4 | 5.1±0.1 | 21.4±2.0 | 2.6±0.4 | 1.1±0.1 | 4.1±0.1 | 18.4±1.4 | 16.8±1.9 | 100±9 |
| 0.225 | 4547 | 16.6±0.2 | 1.94±0.03 | 0.45 | 0.45±0.02 | 3.9±0.2 | 4.7±0.1 | 20.0±1.1 | 2.6±0.2 | 1.1±0.0 | 3.8±0.1 | 16.9±1.0 | 15.7±1.0 | 94±6 |
| 0.27 | 4212 | 18.0±0.1 | 2.18±0.00 | 0.38 | 0.39±0.06 | 2.9±0.4 | 4.6±0.2 | 28.4±4.0 | 1.9±0.4 | 0.8±0.1 | 3.7±0.1 | 15.8±1.3 | 22.7±5.2 | 133±8 |

| $C$ (g/ml) | total time (s) | $c_{NCBD}$ (nM) | $\eta/\eta_0$ | frac. bound (histo) | frac. bound (rates) | $k_{on}$ (s$^{-1}$) | $k_{off}$ (s$^{-1}$) | $K_D$ (nM) | $k_{on,1}$ (s$^{-1}$) | $k_{on,2}$ (s$^{-1}$) | $k_{off,1}$ (s$^{-1}$) | $k_{off,2}$ (s$^{-1}$) | $K_{D,1}$ (nM) | $K_{D,2}$ (nM) |
|---|---|---|---|---|---|---|---|---|---|---|---|---|---|---|
| 0.315 | 3723 | 17.7±0.5 | 2.59±0.02 | 0.37 | 0.38±0.02 | 2.7±0.1 | 4.4±0.1 | 29.4±1.6 | 1.8±0.2 | 0.7±0.0 | 3.7±0.1 | 15.9±1.0 | 23.4±1.9 | 145±5 |
| 0.36 | 3771 | 19.0±0.5 | 3.15±0.09 | 0.35 | 0.36±0.02 | 2.2±0.1 | 4.0±0.1 | 34.4±2.3 | 1.5±0.1 | 0.7±0.1 | 3.4±0.1 | 15.5±2.7 | 26.6±1.5 | 151±14 |

**PEG 200**

| $C$ (g/ml) | total time (s) | $c_{NCBD}$ (nM) | $\eta/\eta_0$ | frac. bound (histo) | frac. bound (rates) | $k_{on}$ (s$^{-1}$) | $k_{off}$ (s$^{-1}$) | $K_D$ (nM) | $k_{on,1}$ (s$^{-1}$) | $k_{on,2}$ (s$^{-1}$) | $k_{off,1}$ (s$^{-1}$) | $k_{off,2}$ (s$^{-1}$) | $K_{D,1}$ (nM) | $K_{D,2}$ (nM) |
|---|---|---|---|---|---|---|---|---|---|---|---|---|---|---|
| 0.045 | 4409 | 15.6±0.3 | 1.13±0.03 | 0.35 | 0.36±0.02 | 3.9±0.2 | 7.0±0.2 | 27.8±1.8 | 2.5±0.2 | 1.0±0.0 | 5.6±0.2 | 22.1±1.6 | 22.2±1.8 | 128±10 |
| 0.09 | 4776 | 16.8±0.4 | 1.34±0.01 | 0.37 | 0.38±0.02 | 3.9±0.2 | 6.5±0.1 | 27.7±1.7 | 2.7±0.2 | 1.0±0.0 | 5.4±0.1 | 21.5±1.0 | 21.9±1.5 | 125±5 |
| 0.135 | 2727 | 16.6±0.3 | 1.52±0.02 | 0.43 | 0.43±0.03 | 4.3±0.2 | 5.7±0.2 | 21.7±1.5 | 2.9±0.1 | 1.2±0.1 | 4.6±0.2 | 19.0±1.3 | 17.2±0.9 | 99±5 |
| 0.18 | 3397 | 17.3±0.2 | 1.75±0.03 | 0.46 | 0.46±0.03 | 4.3±0.1 | 5.0±0.4 | 20.3±1.8 | 2.9±0.3 | 1.1±0.1 | 4.2±0.2 | 17.0±0.9 | 16.0±1.3 | 93±9 |
| 0.225 | 3622 | 15.8±0.3 | 2.06±0.00 | 0.46 | 0.46±0.04 | 3.9±0.3 | 4.6±0.1 | 18.5±1.4 | 2.6±0.4 | 0.9±0.1 | 3.8±0.1 | 15.3±1.3 | 14.8±1.9 | 96±6 |
| 0.27 | 4964 | 15.2±0.4 | 2.47±0.01 | 0.46 | 0.46±0.02 | 2.7±0.1 | 3.2±0.1 | 17.7±1.1 | 1.8±0.1 | 0.7±0.1 | 2.6±0.1 | 12.0±0.7 | 14.0±1.1 | 89±10 |
| 0.315 | 4477 | 17.5±0.7 | 2.88±0.03 | 0.45 | 0.45±0.04 | 2.8±0.2 | 3.4±0.2 | 21.2±2.0 | 1.9±0.3 | 0.7±0.0 | 2.9±0.1 | 11.2±1.3 | 16.9±2.0 | 108±18 |
| 0.36 | 4293 | 16.8±0.5 | 3.53±0.01 | 0.48 | 0.48±0.04 | 2.6±0.2 | 2.9±0.2 | 18.5±1.6 | 1.7±0.2 | 0.6±0.0 | 2.4±0.1 | 10.1±0.6 | 14.8±1.4 | 100±11 |

**PEG 400**

| $C$ (g/ml) | total time (s) | $c_{NCBD}$ (nM) | $\eta/\eta_0$ | frac. bound (histo) | frac. bound (rates) | $k_{on}$ (s$^{-1}$) | $k_{off}$ (s$^{-1}$) | $K_D$ (nM) | $k_{on,1}$ (s$^{-1}$) | $k_{on,2}$ (s$^{-1}$) | $k_{off,1}$ (s$^{-1}$) | $k_{off,2}$ (s$^{-1}$) | $K_{D,1}$ (nM) | $K_{D,2}$ (nM) |
|---|---|---|---|---|---|---|---|---|---|---|---|---|---|---|
| 0.051 | 3059 | 20.2±1.0 | 1.31±0.02 | 0.37 | 0.37±0.03 | 3.9±0.3 | 6.6±0.2 | 34.0±3.2 | 2.5±0.3 | 1.1±0.1 | 5.2±0.2 | 22.2±3.0 | 26.9±2.1 | 148±20 |
| 0.101 | 3049 | 17.5±0.4 | 1.61±0.00 | 0.35 | 0.36±0.04 | 3.0±0.3 | 5.4±0.3 | 31.5±3.3 | 2.0±0.3 | 0.8±0.0 | 4.3±0.1 | 18.9±2.2 | 24.8±2.8 | 144±18 |
| 0.152 | 3404 | 17.8±0.5 | 2.03±0.01 | 0.42 | 0.42±0.03 | 3.0±0.2 | 4.1±0.2 | 24.4±1.9 | 1.9±0.1 | 0.7±0.0 | 3.3±0.1 | 14.1±0.6 | 19.5±0.8 | 127±10 |
| 0.203 | 3063 | 17.8±0.3 | 2.54±0.07 | 0.44 | 0.45±0.03 | 2.7±0.1 | 3.3±0.2 | 22.0±1.6 | 1.7±0.2 | 0.6±0.0 | 2.6±0.1 | 12.1±1.3 | 17.7±1.7 | 122±11 |
| 0.254 | 3090 | 19.9±0.3 | 3.01±0.04 | 0.50 | 0.51±0.04 | 3.0±0.2 | 2.9±0.1 | 19.3±1.7 | 2.0±0.3 | 0.7±0.1 | 2.4±0.1 | 9.7±1.0 | 15.6±1.1 | 105±6 |
| 0.305 | 4161 | 17.2±0.7 | 3.82±0.09 | 0.48 | 0.48±0.04 | 1.8±0.1 | 1.9±0.1 | 18.8±1.5 | 1.1±0.1 | 0.4±0.1 | 1.5±0.0 | 7.4±0.6 | 15.2±1.3 | 113±4 |
| 0.355 | 4391 | 20.1±0.4 | 4.77±0.09 | 0.53 | 0.53±0.07 | 1.9±0.2 | 1.7±0.1 | 17.7±2.1 | 1.2±0.2 | 0.4±0.0 | 1.3±0.0 | 6.2±0.7 | 14.3±1.4 | 112±11 |
| 0.406 | 2798 | 19.0±0.2 | 5.77±0.01 | 0.46 | 0.47±0.06 | 1.2±0.1 | 1.3±0.1 | 21.8±2.8 | 0.7±0.2 | 0.3±0.0 | 1.1±0.1 | 5.2±1.0 | 17.5±3.3 | 127±32 |

**PEG 1000**

| $C$ (g/ml) | total time (s) | $c_{NCBD}$ (nM) | $\eta/\eta_0$ | frac. bound (histo) | frac. bound (rates) | $k_{on}$ (s$^{-1}$) | $k_{off}$ (s$^{-1}$) | $K_D$ (nM) | $k_{on,1}$ (s$^{-1}$) | $k_{on,2}$ (s$^{-1}$) | $k_{off,1}$ (s$^{-1}$) | $k_{off,2}$ (s$^{-1}$) | $K_{D,1}$ (nM) | $K_{D,2}$ (nM) |
|---|---|---|---|---|---|---|---|---|---|---|---|---|---|---|
| 0.045 | 3836 | 15.3±0.2 | 1.32±0.02 | 0.41 | 0.42±0.03 | 4.4±0.2 | 6.2±0.2 | 21.5±1.4 | 3.0±0.2 | 1.4±0.1 | 5.1±0.1 | 21.8±1.1 | 16.7±1.1 | 84±3 |
| 0.09 | 3451 | 16.5±0.6 | 1.69±0.00 | 0.50 | 0.50±0.02 | 5.4±0.2 | 5.3±0.1 | 16.3±0.5 | 3.6±0.1 | 1.7±0.2 | 4.3±0.1 | 20.4±2.1 | 12.6±0.4 | 71±7 |
| 0.135 | 3853 | 17.0±0.5 | 2.18±0.00 | 0.56 | 0.56±0.02 | 4.7±0.1 | 3.7±0.2 | 13.4±0.8 | 3.2±0.1 | 1.6±0.1 | 3.0±0.2 | 14.3±0.8 | 10.4±0.6 | 57±7 |
| 0.18 | 3515 | 16.5±0.5 | 2.85±0.06 | 0.60 | 0.60±0.05 | 5.0±0.3 | 3.3±0.1 | 10.9±0.7 | 3.2±0.3 | 1.5±0.1 | 2.6±0.1 | 12.4±0.9 | 8.5±0.6 | 51±5 |
| 0.225 | 4691 | 15.4±0.3 | 3.64±0.01 | 0.61 | 0.61±0.03 | 3.4±0.2 | 2.2±0.1 | 9.8±0.6 | 2.3±0.1 | 1.1±0.1 | 1.7±0.1 | 9.1±0.6 | 7.5±0.4 | 45±4 |
| 0.27 | 4981 | 15.8±0.4 | 4.91±0.09 | 0.63 | 0.63±0.05 | 2.9±0.2 | 1.8±0.1 | 9.5±0.9 | 1.8±0.2 | 0.7±0.1 | 1.4±0.1 | 6.8±0.4 | 7.6±0.6 | 53±7 |
| 0.315 | 4181 | 14.7±0.7 | 6.23±0.45 | 0.68 | 0.69±0.08 | 2.6±0.3 | 1.2±0.1 | 6.5±0.7 | 1.6±0.1 | 0.8±0.1 | 0.9±0.0 | 4.9±0.2 | 5.1±0.3 | 34±3 |

**PEG 2050**

| $C$ (g/ml) | total time (s) | $c_{NCBD}$ (nM) | $\eta/\eta_0$ | frac. bound (histo) | frac. bound (rates) | $k_{on}$ (s$^{-1}$) | $k_{off}$ (s$^{-1}$) | $K_D$ (nM) | $k_{on,1}$ (s$^{-1}$) | $k_{on,2}$ (s$^{-1}$) | $k_{off,1}$ (s$^{-1}$) | $k_{off,2}$ (s$^{-1}$) | $K_{D,1}$ (nM) | $K_{D,2}$ (nM) |
|---|---|---|---|---|---|---|---|---|---|---|---|---|---|---|
| 0.051 | 3606 | 18.1±0.5 | 1.50±0.08 | 0.46 | 0.46±0.03 | 4.5±0.3 | 5.2±0.3 | 21.1±1.7 | 2.9±0.4 | 1.4±0.2 | 4.1±0.2 | 18.5±1.7 | 16.4±2.3 | 87±9 |
| 0.101 | 4186 | 19.2±0.4 | 2.28±0.12 | 0.45 | 0.46±0.03 | 3.8±0.2 | 4.4±0.2 | 22.4±1.6 | 2.5±0.4 | 1.3±0.1 | 3.6±0.2 | 16.8±1.6 | 17.3±2.2 | 92±10 |
| 0.101 | 2287 | 19.6±0.1 | 2.16±0.10 | 0.46 | 0.46±0.04 | 4.2±0.2 | 4.9±0.5 | 22.9±2.7 | 2.7±0.2 | 1.2±0.2 | 3.9±0.1 | 17.7±1.0 | 18.0±0.9 | 100±8 |
| 0.152 | 3089 | 16.1±0.2 | 3.34±0.11 | 0.52 | 0.52±0.04 | 3.2±0.2 | 2.9±0.1 | 14.7±1.0 | 2.1±0.2 | 0.9±0.1 | 2.3±0.1 | 12.1±1.2 | 11.4±0.7 | 75±11 |
| 0.203 | 3420 | 17.6±0.9 | 4.67±0.37 | 0.60 | 0.60±0.05 | 2.7±0.2 | 1.8±0.1 | 11.8±1.0 | 1.8±0.5 | 0.8±0.1 | 1.5±0.0 | 7.4±0.9 | 9.1±2.1 | 56±6 |
| 0.253 | 3490 | 10.9±0.8 | 6.30±0.00 | 0.50 | 0.50±0.07 | 1.1±0.1 | 1.1±0.1 | 11.0±1.5 | 0.7±0.1 | 0.2±0.0 | 0.9±0.0 | 4.4±0.6 | 8.9±0.8 | 72±6 |

**PEG 4600**

| $C$ (g/ml) | total time (s) | $c_{NCBD}$ (nM) | $\eta/\eta_0$ | frac. bound (histo) | frac. bound (rates) | $k_{on}$ (s$^{-1}$) | $k_{off}$ (s$^{-1}$) | $K_D$ (nM) | $k_{on,1}$ (s$^{-1}$) | $k_{on,2}$ (s$^{-1}$) | $k_{off,1}$ (s$^{-1}$) | $k_{off,2}$ (s$^{-1}$) | $K_{D,1}$ (nM) | $K_{D,2}$ (nM) |
|---|---|---|---|---|---|---|---|---|---|---|---|---|---|---|
| 0.028 | 4803 | 15.8±0.4 | 1.31±0.01 | 0.43 | 0.44±0.02 | 5.3±0.2 | 6.7±0.3 | 20.0±1.3 | 3.4±0.2 | 1.5±0.1 | 5.3±0.1 | 22.5±0.7 | 15.7±0.7 | 86±7 |
| 0.056 | 3249 | 15.3±0.3 | 1.76±0.04 | 0.49 | 0.50±0.01 | 5.8±0.1 | 5.8±0.2 | 15.3±0.6 | 3.8±0.3 | 1.8±0.2 | 4.6±0.3 | 20.5±1.3 | 11.8±1.3 | 62±10 |

| C (g/ml) | total time (s) | $c_{NCBD}$ (nM) | $\eta/\eta_0$ | frac. bound (histo) | frac. bound (rates) | $k_{on}$ (s$^{-1}$) | $k_{off}$ (s$^{-1}$) | $K_D$ (nM) | $k_{on,1}$ (s$^{-1}$) | $k_{on,2}$ (s$^{-1}$) | $k_{off,1}$ (s$^{-1}$) | $k_{off,2}$ (s$^{-1}$) | $K_{D,1}$ (nM) | $K_{D,2}$ (nM) |
|---|---|---|---|---|---|---|---|---|---|---|---|---|---|---|
| 0.084 | 2801 | 14.5±0.3 | 2.47±0.15 | 0.56 | 0.56±0.03 | 5.7±0.2 | 4.5±0.3 | 11.4±0.8 | 3.7±0.3 | 1.9±0.1 | 3.5±0.1 | 16.7±1.3 | 8.8±0.6 | 46±4 |
| 0.112 | 3052 | 13.9±0.1 | 3.08±0.03 | 0.59 | 0.59±0.03 | 5.9±0.2 | 4.1±0.2 | 9.6±0.6 | 3.9±0.2 | 2.1±0.1 | 3.2±0.1 | 16.6±0.9 | 7.3±0.4 | 39±4 |
| 0.141 | 3669 | 14.6±0.3 | 3.90±0.14 | 0.70 | 0.70±0.03 | 6.1±0.2 | 2.6±0.1 | 6.3±0.3 | 4.0±0.3 | 2.3±0.1 | 2.0±0.1 | 9.7±0.7 | 4.8±0.6 | 22±2 |
| 0.169 | 4279 | 14.0±0.9 | 4.76±0.10 | 0.71 | 0.74±0.04 | 6.1±0.3 | 2.2±0.1 | 4.9±0.3 | 3.8±0.3 | 2.4±0.2 | 1.6±0.1 | 7.9±0.6 | 3.7±0.2 | 16±2 |

**PEG 6000**

| C (g/ml) | total time (s) | $c_{NCBD}$ (nM) | $\eta/\eta_0$ | frac. bound (histo) | frac. bound (rates) | $k_{on}$ (s$^{-1}$) | $k_{off}$ (s$^{-1}$) | $K_D$ (nM) | $k_{on,1}$ (s$^{-1}$) | $k_{on,2}$ (s$^{-1}$) | $k_{off,1}$ (s$^{-1}$) | $k_{off,2}$ (s$^{-1}$) | $K_{D,1}$ (nM) | $K_{D,2}$ (nM) |
|---|---|---|---|---|---|---|---|---|---|---|---|---|---|---|
| 0.034 | 3917 | 17.8±0.6 | 1.49±0.00 | 0.48 | 0.48±0.03 | 6.1±0.3 | 6.5±0.3 | 19.1±1.2 | 3.9±0.1 | 1.8±0.1 | 5.1±0.2 | 22.9±1.4 | 14.8±0.6 | 81±7 |
| 0.068 | 2665 | 18.7±0.3 | 2.16±0.02 | 0.55 | 0.56±0.02 | 6.7±0.2 | 5.4±0.2 | 14.9±0.7 | 4.5±0.3 | 2.3±0.1 | 4.3±0.1 | 20.7±0.7 | 11.4±0.7 | 61±4 |
| 0.101 | 3714 | 17.2±0.3 | 3.13±0.02 | 0.65 | 0.65±0.02 | 7.5±0.2 | 4.0±0.1 | 9.1±0.3 | 5.1±0.2 | 2.5±0.1 | 3.2±0.1 | 15.8±0.7 | 6.9±0.3 | 40±2 |
| 0.135 | 3504 | 16.4±1.1 | 4.18±0.07 | 0.68 | 0.69±0.03 | 6.4±0.2 | 2.8±0.1 | 7.2±0.3 | 4.3±0.3 | 2.4±0.1 | 2.2±0.1 | 12.4±2.1 | 5.4±0.3 | 31±7 |
| 0.169 | 2823 | 11.4±0.4 | 5.17±0.24 | 0.68 | 0.69±0.03 | 4.7±0.1 | 2.1±0.1 | 5.1±0.3 | 3.0±0.2 | 2.0±0.2 | 1.5±0.1 | 8.6±0.9 | 3.8±0.4 | 18±4 |

**PEG 35000**

| C (g/ml) | total time (s) | $c_{NCBD}$ (nM) | $\eta/\eta_0$ | frac. bound (histo) | frac. bound (rates) | $k_{on}$ (s$^{-1}$) | $k_{off}$ (s$^{-1}$) | $K_D$ (nM) | $k_{on,1}$ (s$^{-1}$) | $k_{on,2}$ (s$^{-1}$) | $k_{off,1}$ (s$^{-1}$) | $k_{off,2}$ (s$^{-1}$) | $K_{D,1}$ (nM) | $K_{D,2}$ (nM) |
|---|---|---|---|---|---|---|---|---|---|---|---|---|---|---|
| 0.02 | 3461 | 16.4±0.1 | 1.45±0.09 | 0.45 | 0.46±0.02 | 5.5±0.2 | 6.5±0.4 | 19.3±1.4 | 3.7±0.1 | 1.9±0.1 | 5.1±0.2 | 24.8±1.6 | 14.8±0.7 | 78±7 |
| 0.041 | 4633 | 16.9±0.7 | 1.93±0.11 | 0.50 | 0.51±0.04 | 6.6±0.4 | 6.4±0.3 | 16.3±1.3 | 4.4±0.4 | 2.4±0.2 | 5.0±0.2 | 25.1±1.3 | 12.4±1.0 | 65±10 |
| 0.062 | 2818 | 14.6±0.0 | 2.46±0.08 | 0.51 | 0.52±0.04 | 6.0±0.3 | 5.6±0.3 | 13.5±1.0 | 4.0±0.2 | 2.1±0.1 | 4.4±0.2 | 22.8±1.2 | 10.3±0.4 | 56±4 |
| 0.082 | 4258 | 14.8±0.5 | 3.27±0.07 | 0.57 | 0.58±0.05 | 6.2±0.4 | 4.6±0.4 | 10.9±1.2 | 4.1±0.4 | 2.2±0.1 | 3.6±0.1 | 18.7±0.5 | 8.3±0.7 | 46±3 |
| 0.102 | 3084 | 13.2±0.8 | 3.97±0.17 | 0.56 | 0.56±0.01 | 5.4±0.1 | 4.2±0.2 | 10.3±0.4 | 3.5±0.1 | 1.9±0.1 | 3.2±0.1 | 18.4±1.8 | 7.8±0.3 | 46±6 |
| 0.123 | 3006 | 10.3±0.2 | 5.02±0.32 | 0.62 | 0.65±0.02 | 6.4±0.1 | 3.4±0.1 | 5.5±0.2 | 4.1±0.1 | 2.3±0.1 | 2.6±0.0 | 15.5±1.5 | 4.2±0.1 | 25±1 |